\documentclass{jpp}
\usepackage{graphicx}
\usepackage{subcaption}
\usepackage[utf8]{inputenc}
\usepackage[T1]{fontenc}
\usepackage{amsmath}

\shorttitle{Plasmoid formation via competing cross-scale instabilities}
\shortauthor{S. Thatikonda, F. N. De Oliveira-Lopes, A. Mustonen, K. Pommois, D. Told and F. Jenko}

\title{Plasmoid formation via competing lower-hybrid drift and Kelvin--Helmholtz instabilities: A hybrid kinetic--gyrokinetic simulation study}

\author{S. Thatikonda\aff{1}
  \corresp{\email{sreenivasa.thatikonda@ipp.mpg.de}},
  F. N. De Oliveira-Lopes\aff{1,2},
 A. Mustonen\aff{1},  K. Pommois\aff{1},  D. Told\aff{1} \and F. Jenko\aff{1}}

\affiliation{\aff{1}Max-Planck Institute for Plasma Physics, Boltzmannstrasse 2, 
            Garching,
            85748, 
            Bavaria,
            Germany
\aff{2}Centre for mathematical Plasma Astrophysics, KU Leuven, Celestijnenlaan 200B, Leuven, Belgium}

\begin{document}

\maketitle

\begin{abstract}
We investigate the nonlinear formation of plasmoids in 2D low-$\beta_e$ current sheets through the interplay between the Kelvin–Helmholtz instability (KHI) and the lower-hybrid drift instability (LHDI). Using a hybrid kinetic--gyrokinetic model-based \textit{Super Simple Vlasov} (ssV) code with fully kinetic ions and drift-kinetic electrons, we simulate Harris-type current sheets and velocity shear layers with strong cross-field density gradients. Our central hypothesis is that steep density gradients drive LHDI, which can grow faster than KHI and initiate an inverse cascade from kinetic to fluid scales, potentially suppressing KHI. Our simulations confirm that, in thin current sheets, LHDI develops rapidly at the sheet edges and nonlinearly merges into larger-scale magnetic islands before KHI can evolve. These LHDI-driven structures distort the velocity shear and suppress classical KH vortices. In contrast, for thicker current sheets or weaker density gradients, KHI dominates and produces the expected rolled-up vortices and associated plasmoids. These findings demonstrate that LHDI-induced turbulence can act as both a seed and a regulator of plasmoid-generating instabilities, mediating cross-scale energy transfer. This mechanism is relevant to thin boundary layers in space plasmas, such as the solar wind magnetosphere interface, and suggests that microturbulence can govern large-scale magnetic topology during collisionless reconnection.
\end{abstract}
\section{Introduction}

Magnetic reconnection in thin current sheets often produces plasmoid chains (magnetic islands or flux ropes) as a consequence of tearing instabilities and secondary island merging~\citep{Loureiro2007,Samtaney2009,Comisso2017}. In high-Lundquist-number plasmas, a cascade of plasmoid formation is well-known to occur, significantly enhancing reconnection rates~\citep{Loureiro2007,Bhattacharjee2009,Huang2010}. However, in addition to the standard MHD tearing modes, there is growing evidence that microscopic plasma instabilities in the current sheet can influence the plasmoid formation process~\citep{Ricci2004,Kowal2019,Borgogno2022}. In particular, two candidates are the \textit{lower-hybrid drift instability} (LHDI) and the \textit{Kelvin--Helmholtz instability} (KHI)~\citep{Daughton2003,Fujimoto1994,Yuan2024}. This paper explores how these two instabilities, operating on disparate scales, nonlinearly interact to affect plasmoid generation in 2D current sheets.

The lower-hybrid drift instability (LHDI) is a cross-field instability driven by sharp density gradients and associated diamagnetic drifts. It typically excites fluctuations at frequencies near the lower-hybrid frequency $\omega_{LH}$ and at wavelengths on the order of the ion inertial length or smaller~\citep{Daughton2003,Davidson1977,Gary1979}.
Early studies showed that short-wavelength LHDI modes (comparable to the electron gyroradius) can grow rapidly on the flanks of thin current sheets where the density gradient is steep. These modes are usually quasi-electrostatic and localized to the sheet edges~\citep{Daughton2003,Huba1983,Gary1993}. The lower-hybrid drift instability (LHDI) can drive turbulence, enhance cross-field transport, and contribute to anomalous resistivity in thin current sheets~\citep{Daughton2005,Huba1980,Cattell2005}.
Lower-hybrid turbulence has been directly observed in laboratory reconnection experiments—such as the MRX device—and in space plasmas, where current layers at the magnetopause and magnetotail frequently exhibit lower-hybrid wave activity during reconnection~\citep{Yoo2020,Ren2005,Carter2001,Tang2013,Graham2017}. Historically, it was debated whether the lower-hybrid drift instability (LHDI) significantly alters magnetic reconnection or is merely a byproduct. Recent kinetic simulations indicate that while LHDI alone may not dramatically change the global reconnection rate, it does restructure the current sheet and can seed secondary instabilities. For instance, LHDI-driven edge turbulence may initiate downstream modes such as the drift-kink or Kelvin–Helmholtz instabilities, thereby indirectly promoting magnetic island formation~\citep{Roytershteyn2012,Roytershteyn2013,Baumjohann2010}.

The KHI, on the other hand, is a fluid-like instability of shear flows, producing vortical structures (Kelvin--Helmholtz vortices) at velocity discontinuities~\citep{Fujimoto1994}. In magnetized plasmas, the Kelvin–Helmholtz instability (KHI) can arise at boundary layers with strong velocity shears, such as the flanks of magnetopauses or in reconnection outflows~\citep{Borgogno2022,Fujimoto1994,Hasegawa2004}.
 Fully rolled-up KHI vortices have been observed (e.g., at the Earth’s magnetopause) and are known to entrain plasma from one side to the other, facilitating plasma mixing across what would otherwise be closed boundaries. In some cases, the internal shear of KHI vortices can induce magnetic reconnection within them (so-called vortex-induced reconnection), leading to the formation of flux ropes inside KH waves. Thus, KHI can generate plasmoid-like magnetic islands as well, albeit through a different mechanism than the standard tearing mode~\citep{Borgogno2022,Hasegawa2004}. The KHI typically acts on larger spatial scales (comparable to the layer thickness or larger) and slower growth rates than LHDI, 
because it is driven by macroscopic flow shear, whereas the LHDI arises from cross-field microscopic drifts associated with velocity shear due to density gradients~\citep{Borgogno2022,Daughton2003}.

Recent works have begun to examine interactions between LHDI and KHI in boundary layers. Notably,~\citet{Dargent2019} performed fully kinetic particle-in-cell simulations of a 
layer featuring both macroscopic velocity shear and strong cross-field density gradients (conditions analogous to Mercury’s magnetopause), 
and found that LHDI and KHI can \textit{coexist and compete}.
 They showed that depending on the density gradient, LHDI can even dominate the dynamics: because LHDI grows on much shorter time and length scales, it can saturate nonlinearly before KHI gets well underway. In such cases, the LHDI generates an inverse cascade—small-scale kinetic eddies merge into larger ones—producing structures at the scale where KHI would normally form. These LHDI-induced meso-scale structures effectively preempt the KHI by smoothing out or redistributing the velocity shear, thereby suppressing the standard KHI vortices. On the other hand, if the density gradient is weaker (or the shear flow stronger), the KHI can grow relatively unimpeded and may overshadow the LHDI.~\citet{Dargent2019} concluded that the outcome depends on a combination of the density gradient strength, velocity shear, and layer thickness. Their simulations suggested that at Mercury’s magnetopause—characterized by extreme density contrast and low plasma beta—the LHDI likely dominates plasma mixing. In contrast, at Earth’s magnetopause, where density gradients are more moderate, the KHI is typically the primary instability, with LHDI playing a secondary role~\citep{Dargent2019}.

Additional evidence of LHDI–KHI interplay comes from both simulations and observations. For instance, recent two-fluid simulations by~\citet{Yuan2024} showed that during the nonlinear stage of KHI at Earth’s magnetopause, bursty LHDI activity appears around the edges of Kelvin–Helmholtz vortices and diminishes as the vortices saturate. This implies a bidirectional coupling: not only can LHDI seed or modify KHI, but the KHI vortices can in turn generate localized density gradients within the rolled-up flow that trigger LHDI~\citep{Norgren2012,Wilder2019}. Thus, the relationship between these instabilities is complex and highly context-dependent. 

In collisionless magnetotail experiments, LHDI has been observed to accompany secondary instabilities such as the kink mode during current sheet thinning~\citep{Graham2017}. On the dayside magnetopause, lower-hybrid waves often co-occur with Kelvin–Helmholtz waves~\citep{Tang2013,Hasegawa2004}, further supporting the notion of interdependent instability dynamics in boundary layers.

Plasmoid formation in magnetic reconnection is typically attributed to the tearing instability, which generates multiple X-points and O-points in a thinning current sheet~\citep{Loureiro2007,Bhattacharjee2009}. However, when microinstabilities such as the lower-hybrid drift instability (LHDI) and the Kelvin–Helmholtz instability (KHI) are present, they can significantly modify how plasmoids nucleate and evolve~\citep{Roytershteyn2013,Dargent2019}. LHDI-induced turbulence can fragment the current sheet at kinetic scales, potentially “seeding” plasmoids or allowing small flux ropes to form from thermal or numerical noise~\citep{Daughton2003,Yoo2020}. If KHI develops at reconnection outflow boundaries, it can roll up the outflow jet and entrain magnetic flux, creating large-scale vortices that may later undergo secondary reconnection~\citep{Borgogno2022,Hasegawa2004}.

These processes introduce multiscale turbulence into an otherwise orderly plasmoid chain produced by linear tearing. Kinetic simulations of asymmetric reconnection—with realistic magnetopause-like density gradients—have shown that while LHDI often emerges near current sheet boundaries, it does not necessarily enhance the global reconnection rate~\citep{Roytershteyn2012,Roytershteyn2013}. Instead, its primary role is to broaden the current sheet and promote cross-field particle mixing, which can influence plasmoid coalescence or lifetime~\citep{Daughton2003,Graham2017}. Similarly, KHI has been proposed as a trigger for reconnection at the flanks of current sheets—especially in the presence of strong outflow shear—highlighting that secondary instabilities can initiate plasmoid formation through mechanisms distinct from classical tearing~\citep{Borgogno2022,Fujimoto1994}.

In this paper, we aim to systematically study the nonlinear interplay of LHDI and KHI in plasmoid-producing current sheets, using a hybrid kinetic--gyrokinetic simulation approach. By treating ions kinetically and electrons with a drift-kinetic model, we capture the relevant physics of LHDI (which requires resolving electron drifts and Hall effects) while also allowing long-timescale evolution of large structures like KHI and plasmoids. We focus on low-$\beta_e$, solar-wind-like plasma conditions. The simulations include two setups: (1) a Harris-type current sheet that undergoes magnetic reconnection and plasmoid formation, and (2) an isolated velocity shear layer (with co-existing magnetic and density gradients) prone to KHI. In both cases, a strong cross-field density gradient is present to drive LHDI.

We focus on how parameters such as the current-sheet half-thickness, the ion--electron
mass ratio $m_i/m_e$, the plasma beta $\beta_e$, and the ion to electron temperature ratio
influence the relative strength of LHDI and its ability to modify the current-sheet
structure. A detailed parametric investigation of LHDI itself has been performed in our
previous work~\citep{Thatikonda2025}, where its dependence
on mass ratio, temperature ratio, $\beta$, and gradient scale lengths was quantified.
Here, those results provide the foundation for interpreting how LHDI-driven
restructuring impacts the subsequent development of larger-scale dynamics. Although the
present study does not include a systematic parameter scan of KHI, the comparison among
different shear-layer configurations enables us to distinguish cases in which LHDI
saturates early and drives plasmoid formation through an inverse cascade from cases in
which the background shear is sufficiently strong for KHI or tearing to dominate the
formation of magnetic islands.

The remainder of the paper is organized as follows. In Section~\ref{sec:model}, we describe the simulation model and initial conditions in detail. Section~\ref{sec:linear} presents the linear evolution of instabilities in our simulations. Section~\ref{sec:nonlinear} details the fully nonlinear evolution and plasmoid formation. Section~\ref{sec:discussion} discusses the implications of our results and comparisons with past observations and simulations. Finally, Section~\ref{sec:conclusion} summarizes our conclusions and suggests directions for future research.

\section{Numerical Model and Simulation Setup}
\label{sec:model}

We perform two-dimensional simulations using a hybrid kinetic--gyrokinetic code based on a semi-Lagrangian Vlasov formulation (ssV), where ions are treated fully kinetically via the Vlasov equation and electrons are modeled using a drift-kinetic approximation~\citep{your2025paper}. This hybrid approach enables accurate capture of ion-scale dynamics and the lower-hybrid drift instability (LHDI) physics while keeping computational costs manageable.
\subsection{Model Description}

The ion dynamics are governed by the full Vlasov equation for the ion distribution function $f_i(\mathbf{R}_i, \mathbf{v}_i, t)$:
\begin{equation}
\frac{\partial f_i}{\partial t} + \mathbf{v}_i \cdot \nabla f_i + \frac{q_i}{m_i} \left( \mathbf{E} + \mathbf{v}_i \times \mathbf{B} \right) \cdot \nabla_{\mathbf{v}} f_i = 0,
\end{equation}
where $\mathbf{E}$ and $\mathbf{B}$ are the total electric and magnetic fields.

Electrons are treated using a drift-kinetic model with distribution function $f_e(\mathbf{R}_e, v_\parallel, \mu, t)$, where $\mu = m_e v_\perp^2 / (2B)$ is the magnetic moment and $v_\parallel$ is the parallel velocity. The electron evolution is described by:
\begin{align}
\frac{\partial f_e}{\partial t}
&+ \left( v_{gy,\parallel} + \frac{e}{m} A_{1\parallel} \right) \cdot \nabla_{gy} f_e + \frac{\hat{b}}{eB_0} \times 
\left( e \nabla \phi_1 - e v_{gy,\parallel} \nabla A_{1\parallel} \right) \cdot \nabla_{gy} f_e \nonumber \\
&- \left( \frac{e}{m} \nabla \phi_1 - \frac{e}{m} v_{gy,\parallel} \nabla A_{1\parallel} \right) 
\cdot \frac{\partial f_e}{\partial v_{gy,\parallel}} = 0
\end{align}
In what follows, the subscript “gy” refers to gyrocenter quantities. The parallel direction, indicated by the symbol $\parallel$, is defined relative to the equilibrium magnetic field $\mathbf{B}_0$. Accordingly, $v_{\mathrm{gy},\parallel}$ represents the gyrocenter velocity
component parallel to $\mathbf{B}_0$, and $\nabla_{\mathrm{gy}}$ denotes spatial gradients
evaluated at the gyrocenter position.
Quasi-neutrality is imposed via:
\begin{equation}
n_i = n_e,
\end{equation}
with $n_i = \int f_i \, d^3v$ for ions and $n_e = \int f_e \, dv_\parallel$.

The self-consistent fields are obtained by solving coupled Poisson and Ampère equations. 
The generalized Poisson equation is
\begin{equation} \label{eq:poissoneq}
\frac{1}{4\pi}\nabla_{\perp}^{2}\phi_{1}
\left(4\pi\frac{\rho_{th}^{2}}{\lambda_{D}^{2}} - 1\right)
+ u_{e \parallel} \frac{\rho_{th}^{2}}{\lambda_{D}^{2}} \nabla_{\perp}^2 A_{1 \parallel}(\mathbf{x})
= \sum_i q_i n_i(\mathbf{x}) + q_e\,n_e(\mathbf{x}),
\end{equation}
where $\rho_{th}$ is the ion thermal Larmor radius, $\lambda_D$ is the Debye length, 
and $u_{e\parallel}$ is the parallel electron flow velocity. Here $n_i(\mathbf{x})$ and $n_e(\mathbf{x})$ are the ion and electron number densities, 
$q_i$ is the charge of ion species $i$ (with $q_i = Z_i e$ and $e>0$ the elementary charge), 
and the electron charge is $q_e = -e$.

The parallel component of the vector potential $A_{1\parallel}$ is advanced using the coupled Ampère equation:
\begin{equation} \label{eq:Ampereeq}
\frac{c}{4\pi}\nabla_{\perp}^{2} A_{1\parallel}\left(1+\frac{\beta_e}{2}\right)
+ u_{e \parallel}\frac{\rho_e^2}{\lambda_D^2}\nabla_{\perp}^2\phi_1(\mathbf{x})
= \frac{e^2}{m_e c} n_e A_{1 \parallel}(\mathbf{x}) + I_e + \sum_i I_{i \parallel},
\end{equation}
where $\rho_e$ is the electron Larmor radius, $\beta_e$ is the electron beta, 
and $I_e=q_e\,n_e(\mathbf{x})\,u_{e\parallel}(\mathbf{x})$ and $I_{i\parallel}=q_i\,n_i(\mathbf{x})\,u_{i\parallel}(\mathbf{x})$ are the electron and ion parallel current contributions, respectively.

The magnetic field is split as $\mathbf{B} = \mathbf{B}_0 + \delta \mathbf{B}$, where $\delta \mathbf{B} = \nabla \times \mathbf{A}$ and $\mathbf{B}_0$ is the equilibrium guide field. Hall physics and finite Larmor radius effects are fully captured through ion kinetics, while electrons contribute parallel current and support the equilibrium through their drifts.




\subsection{Normalization and Geometry}

The simulations are carried out in slab geometry in the $(x, y)$ plane with a background magnetic field (guide field) along the $z$-direction. 

Normalization is as follows:

In this work, all reference quantities denoted by the subscript $R$ are defined using ion parameters; specifically, the reference mass, charge, cyclotron frequency, and Larmor radius correspond to the ion species.
Time and lengths are normalized using a reference cyclotron frequency ($\Omega_{cR}$) and radius ($\rho_R$) as $t = \Omega_{cR}^{-1} \hat{t}$ and $d = \rho_R \hat{d}$, where $\Omega_{cR} = \frac{q_R B_R}{m_R c}$ and $\rho_R = \frac{m_R c_R}{q_R B_R}$. Here, $c_R$ is the reference thermal velocity, which is calculated as $c_R = \sqrt{\frac{T_R}{m_R}}$. Velocity is normalized using a reference thermal velocity $c_R$ as
$v = c_R \hat{v} \hat{v}_{T\sigma}$, where $v_{T\sigma} = \sqrt{2 T_\sigma / m_\sigma}$
is the thermal velocity of species $\sigma$, and
$\hat{v}_{T\sigma} = v_{T\sigma}/c_R$ is its dimensionless form.
Here, $\hat{v}$ denotes the normalized velocity coordinate.

We consider two physical setups:
\begin{enumerate}
  \item A Harris-type current sheet with reversed magnetic field and density profile centered at $x = 0$.
  \item A shear layer with imposed velocity $V_y(x)$ and co-located density gradient to allow LHDI and KHI growth.
\end{enumerate}

Both configurations maintain pressure balance. The domain uses periodic boundary conditions in $x$ and $y$.

\subsection{Initial Conditions}

The Harris sheet equilibrium is given by:
\begin{align}
B_y(x) &= B_0 \tanh(x/L), \\
n(x) &= n_0 \,\text{sech}^2(x/L),
\end{align}
with corresponding electron drift velocity $v_{de} = -j(x)/(en)$ ensuring current balance, where \(L\) is the half–thickness of the current sheet.

The shear flow setup employs a hyperbolic tangent profile:
\begin{align}
V_y(x) &= V_0 \tanh(x/L), \\
n(x) &= n_0 \left(1 - \Delta_n \tanh(x/L)\right),
\end{align}
allowing density and velocity gradients to co-exist.

\begin{figure}
\centering
\includegraphics[width=0.48\textwidth]{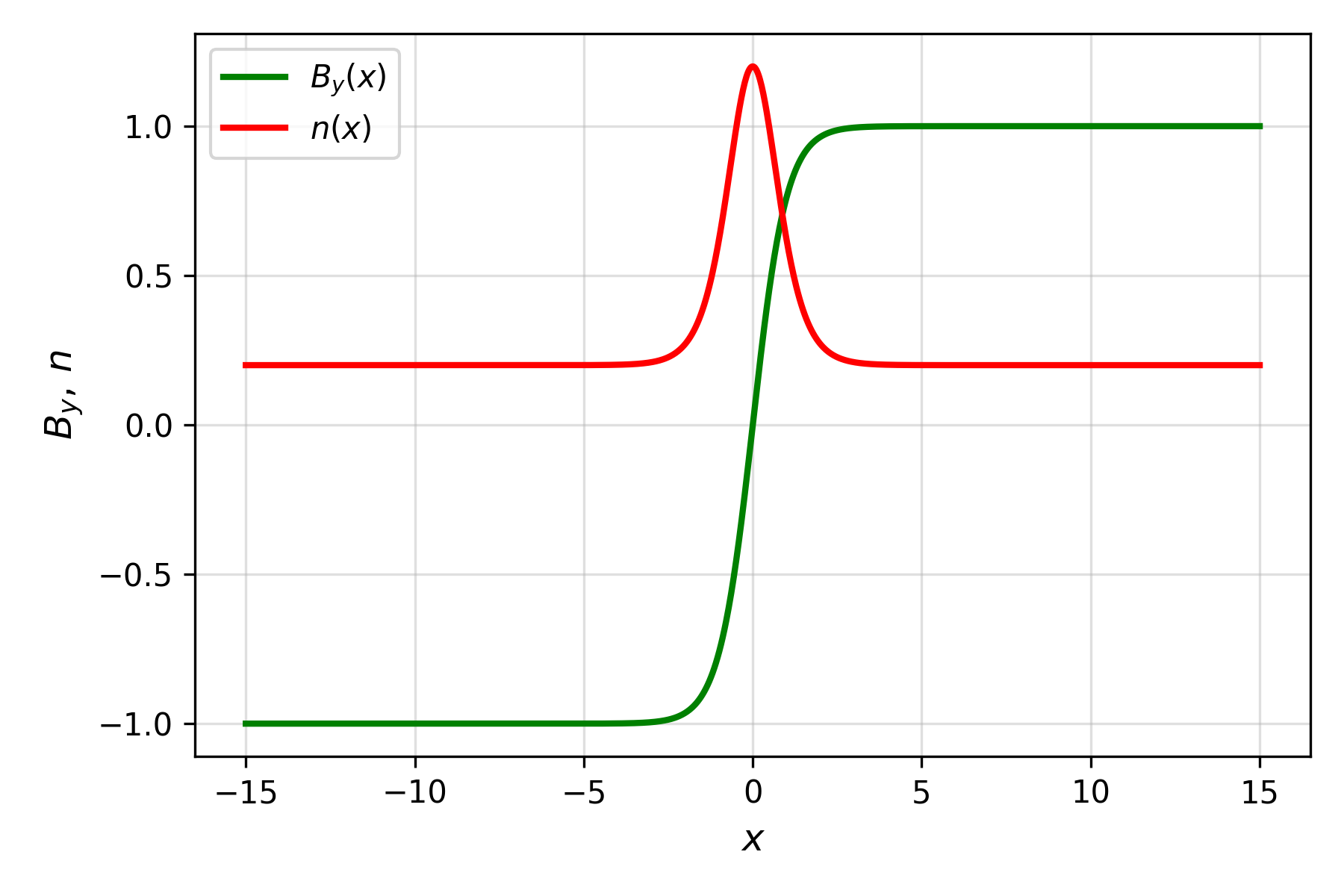}\hfill
\includegraphics[width=0.48\textwidth]{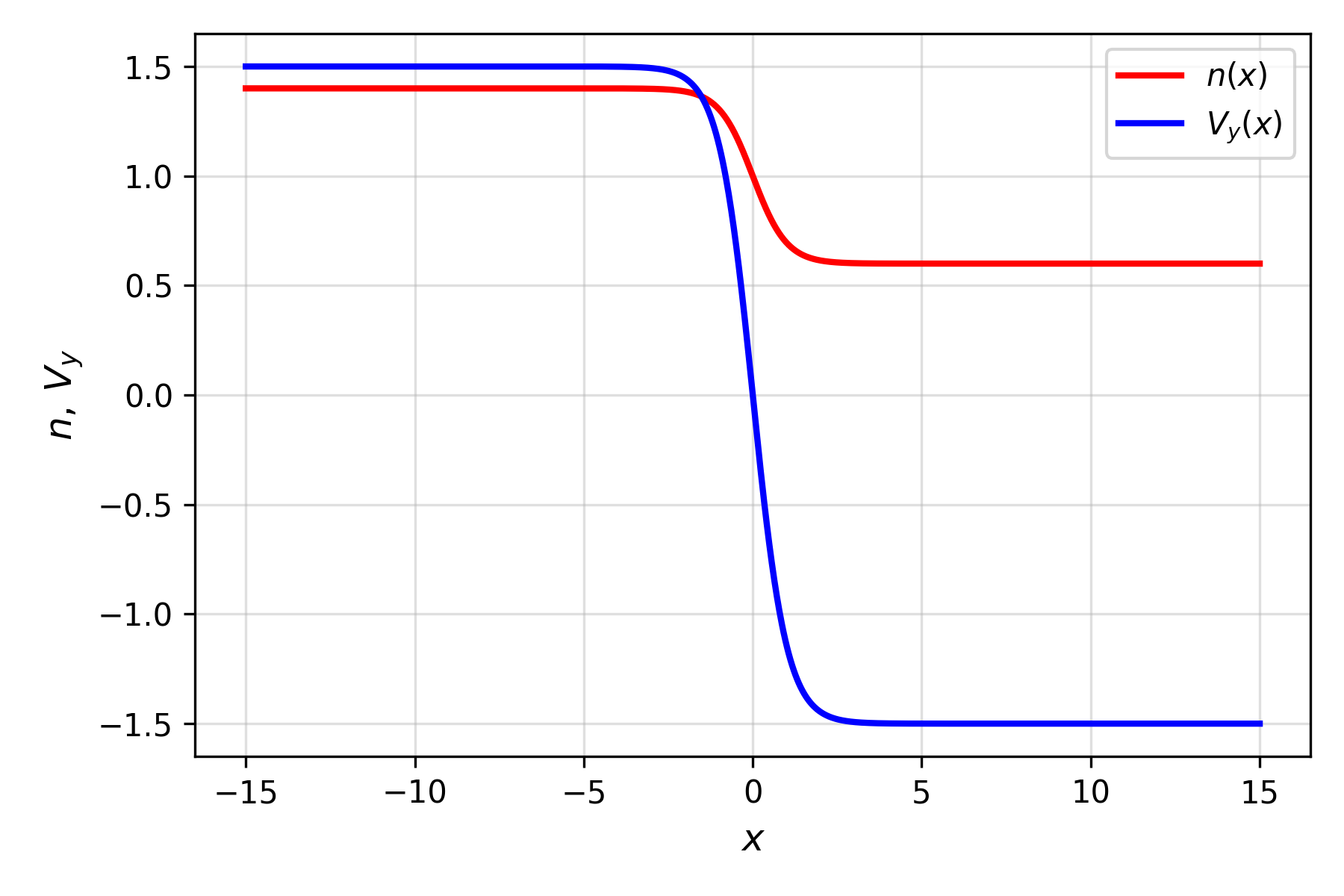}
\caption{Initial equilibrium profiles for (left) the Harris-type current sheet and (right) the shear-layer configuration, showing $n(x)$, $B_y(x)$, and $V_y(x)$ where applicable.}
\label{fig:init_profiles}
\end{figure}

The ion distribution function is initialized as:
\begin{equation}
f_i(x, y, v_x, v_y) = n(x) \cdot \frac{1}{2\pi v_{th,i}^2} \exp\left(-\frac{v_x^2 + (v_y-V_y(x))^2}{2 v_{th,i}^2} \right),
\end{equation}
and the electron distribution follows an analogous drift-kinetic Maxwellian profile. The initial equilibrium profiles for Harris current sheet and shear layer cases were shown in Fig~\ref{fig:init_profiles}. Small perturbations are applied to the ion distribution function through a stream function $\delta\psi(x,y)$ defined in the perpendicular $(x,y)$ plane for both equilibrium configurations.
 For the Harris
current sheet, the lower–hybrid drift instability is seeded using a two–dimensional
mode,
\[
\delta\psi_{\mathrm{LHDI}}(x,y)
=\psi_0
\cos\!\left(\frac{2\pi x}{L_x}\right)
\cos\!\left(\frac{2\pi y}{L_y}\right),
\]
For the shear--layer configuration, the perturbation is localized across the layer
and takes the form
\[
\delta\psi_{\mathrm{shear}}(x,y)
=\psi_0
\exp\!\left[-\tfrac{1}{2}\left(\tfrac{y}{L_{y}}\right)^2\right]
\sin\!\left(\frac{2\pi x}{L_x}\right),
\]
with $\psi_0 = 10^{-3}$ and $L_x$ and $L_y$ denote the domain lengths in the $x$ and $y$ directions,
respectively. Different perturbations are used because LHDI in the Harris sheet responds to
two–dimensional edge fluctuations, whereas the shear layer requires a streamwise
modulation to seed KHI efficiently.

\subsection{Simulation Parameters}

The following Table~\ref{tab:params} summarizes the fiducial simulation parameters used throughout this work, where $\,v_{i,\max}, \,v_{e,\max}$ denote the half-width of the velocity domain in phase space for each species:

\begin{table}
    \begin{center}
\def~{\hphantom{0}}
\begin{tabular}{lccc}
Parameter & Symbol & Value \\[3pt]
Ion-to-electron mass ratio & $m_i/m_e$ & 36 \\
Ion-to-electron temp. ratio & $T_i/T_e$ & 10.0 \\
Electron plasma beta & $\beta_e$ & 0.01 \\
Current sheet half-width & $L$ & $1.0\,\rho_i$ \\
Domain size & $(L_x, L_y)$ & $(12.8\,\rho_i, 6.4\,\rho_i)$ \\
Grid resolution & $\Delta x, \Delta y$ & $0.025\,\rho_i$ \\
Time step & $\Delta t$ & $0.005\,\Omega_{ci}^{-1}$ \\
Ion velocity box size & $[-v_{i,\max},\,v_{i,\max}]$ & $[-5\,v_{th,i},\,5\,v_{th,i}]$ \\
Electron velocity box size & $[-v_{e,\max},\,v_{e,\max}]$ & $[-5\,v_{th,e},\,5\,v_{th,e}]$ \\
Initial perturbation & $\psi_0$ & $10^{-3} $\\
Guide field & $B_0$ & 1.0\\
\end{tabular}
\caption{Simulation parameters.}
\label{tab:params}
\end{center}
\end{table}

\begin{figure}
\centering
\includegraphics[width=0.5\textwidth]{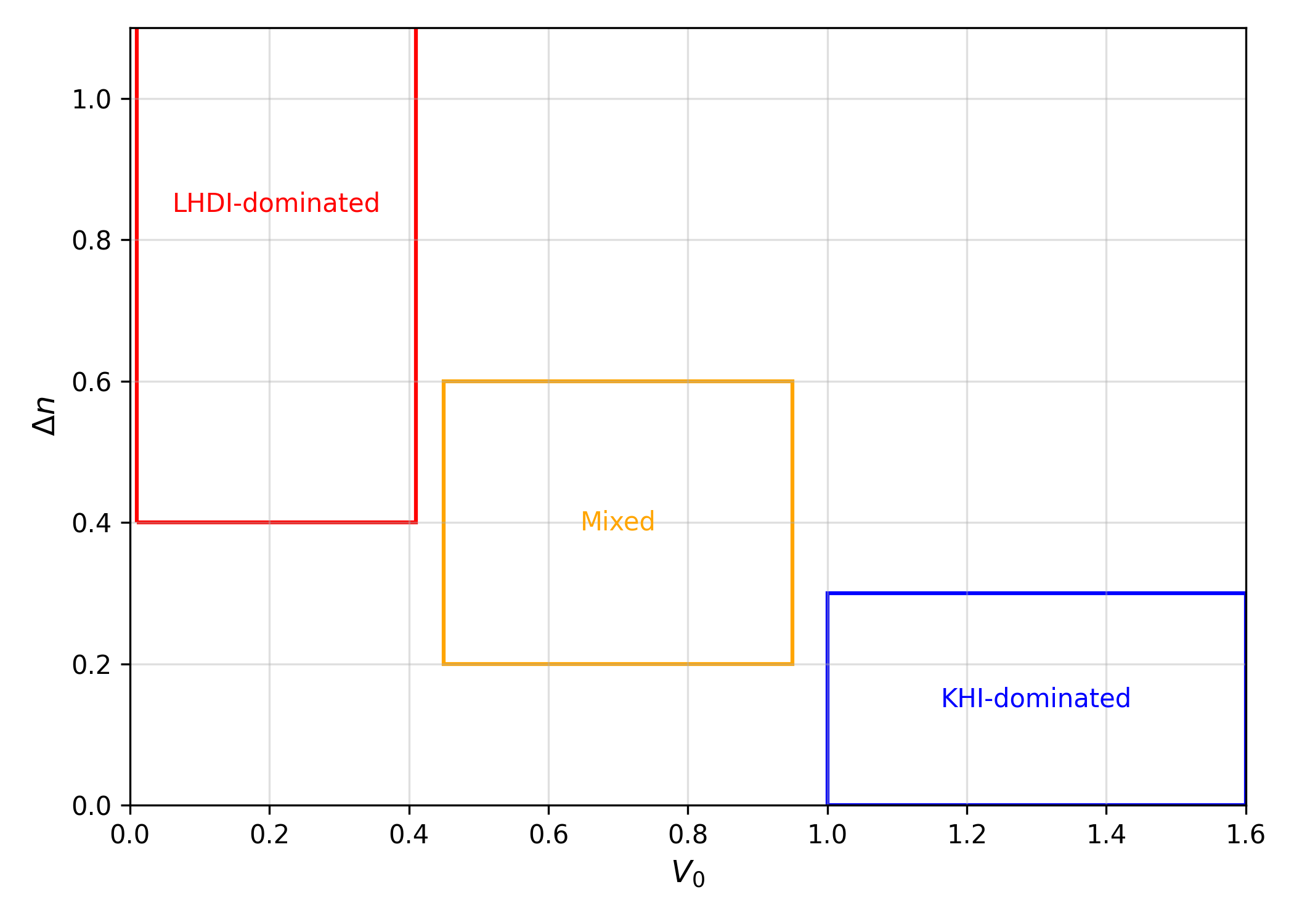}
\caption{Qualitative parameter–space sketch showing the LHDI, mixed, and KHI
regimes associated with the shear–layer runs.}
\label{fig:V0_deltan_map}
\end{figure}

The parametric dependence of LHDI on mass ratio, temperature ratio, plasma beta, and layer thickness was examined in our earlier study~\citep{Thatikonda2025}, and those results will be referenced in the discussion. In the present work these parameters are held fixed as listed in Table~\ref{tab:params}; instead, we investigate four distinct configurations: (i) a Harris sheet exhibiting LHDI, (ii) a shear layer in which LHDI is dominant, (iii) a mixed shear layer where LHDI and KHI grow concurrently, and (iv) a shear layer where the velocity shear is large enough for KHI to dominate the early evolution. For the shear–layer configurations we vary the shear amplitude $V_0$ and density contrast $\Delta n$ in order to obtain three distinct regimes while keeping all other physical parameters fixed (Table~\ref{tab:params}). The LHDI–dominated case uses a strong density gradient and weak shear, $(V_0,\Delta n) = (0.25,\,1.0)$. The mixed regime employs comparable gradients with $(V_0,\Delta n) = (0.5,\,0.5)$. The KHI–dominated case is obtained by increasing the velocity shear and reducing the density contrast, $(V_0,\Delta n) = (1.5,\,0.25)$. These choices position the three runs in distinct regions of the qualitative
$(V_0,\Delta n)$ parameter space, as illustrated in Figure~\ref{fig:V0_deltan_map}.

 Runs were typically evolved for $t = 200$--$400\,\Omega_{ci}^{-1}$ with diagnostics output at intervals of $1.0\,\Omega_{ci}^{-1}$.

\section{Linear Instability Growth and Mode Competition}
\label{sec:linear}
\subsection{Linear evolution in Harris sheet configuration}
During the early linear phase, the Harris current sheet supports the rapid growth of the
lower–hybrid drift instability (LHDI) driven by the strong density gradient at the sheet
edges. The instability first appears as quasi–electrostatic fluctuations localized on
either side of the current layer, with wave vectors predominantly oriented in the
$y$--direction. These modes exhibit electron–scale perpendicular structure
($k_\perp \rho_e \sim 1$) and grow rapidly under the combined influence of the local
density gradient and diamagnetic drifts.

\begin{figure}
\centering
\begin{subfigure}[t]{0.49\textwidth}
    \centering
    \includegraphics[width=\textwidth]{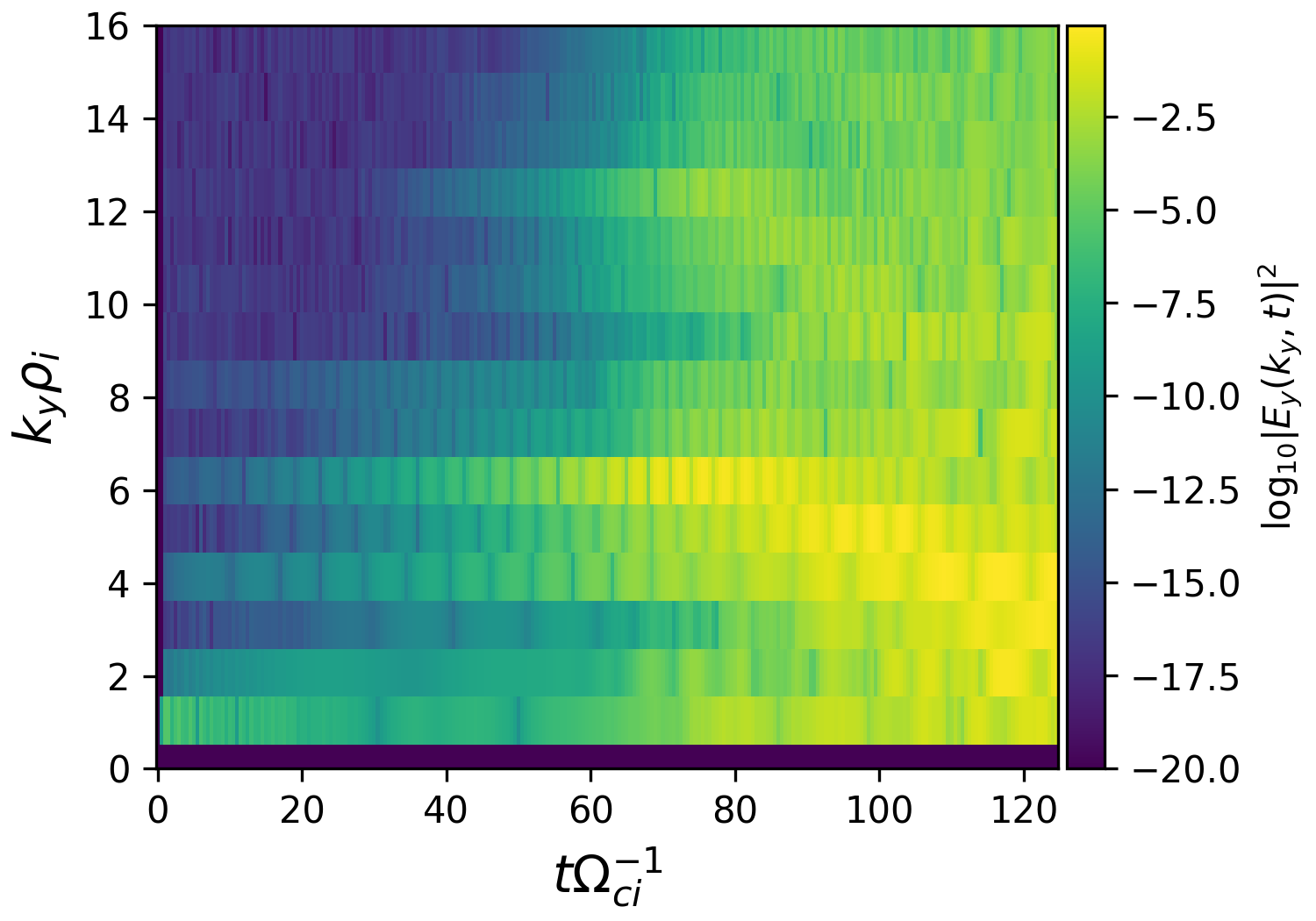}
    
\end{subfigure}
\hfill
\begin{subfigure}[t]{0.49\textwidth}
    \centering
    \includegraphics[width=\textwidth]{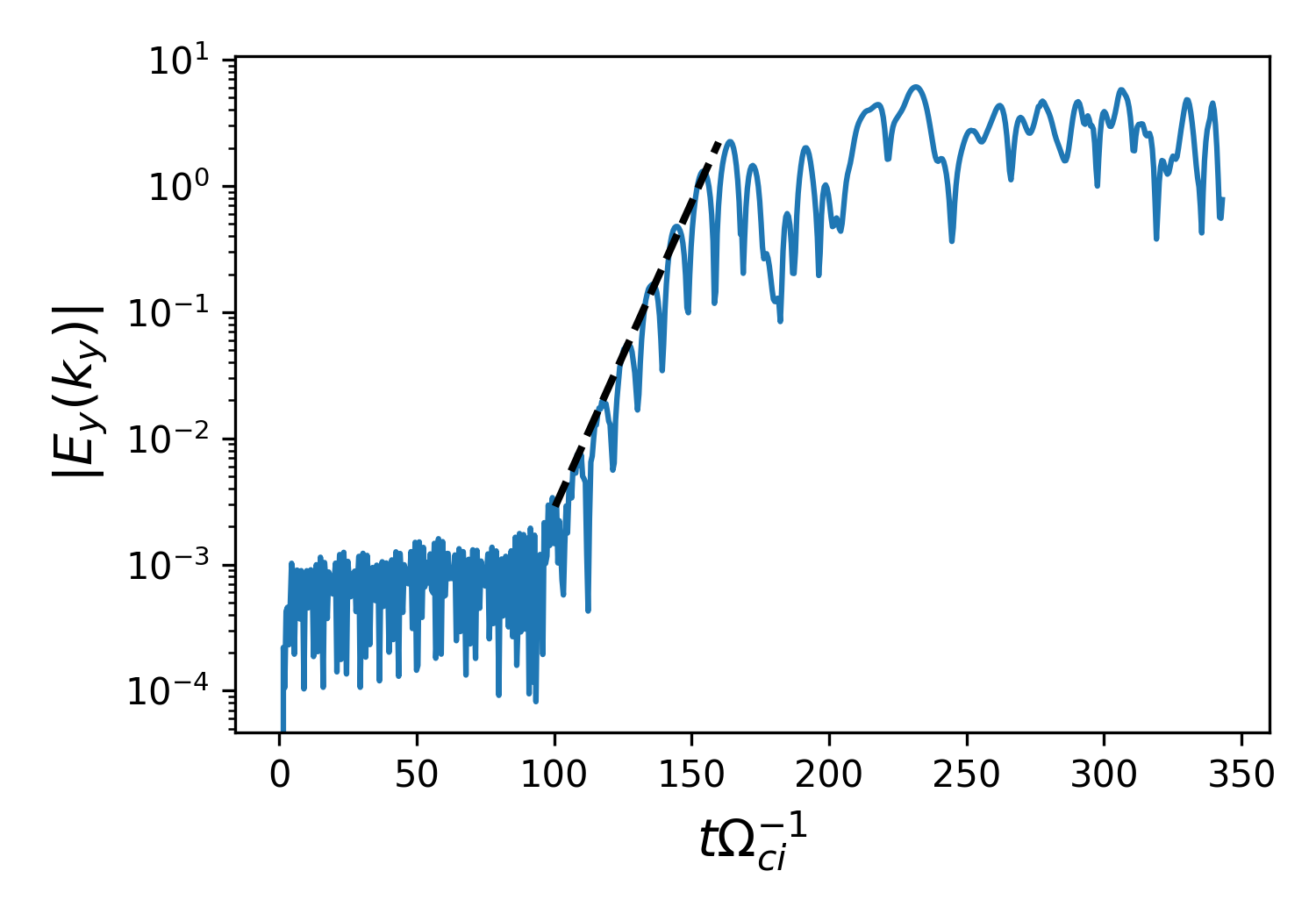}

\end{subfigure}

\caption{Linear evolution of the lower–hybrid drift instability in the Harris current sheet for $m_i/m_e=36$, $T_i/T_e=10$, and $L=1\rho_i$: (a) electric–field spectral energy showing a dominant peak near $k_y\rho_i\approx 6$, and (b) corresponding temporal growth rate.}
\label{fig:harris_spectrum_growth}
\end{figure}

Figure~\ref{fig:harris_spectrum_growth}(a) shows the corresponding electric--field
spectrum, where a clear enhancement of $E_{E_y}(k_y)$ appears around 
$k_y \rho_i \approx 6$, consistent with the characteristic LHDI range for the mass
ratio used here ($m_i/m_e = 36$). The growth rate is evaluated from the temporal
evolution of Fourier amplitudes,
\[
\gamma(k) = \frac{1}{\Delta t}
\log\!\left(
\frac{|\hat{E}_y(k,t+\Delta t)|}{|\hat{E}_y(k,t)|}
\right),
\]
where $\hat{E}_y(k,t)$ denotes the Fourier-transformed electric field. As illustrated
in Figure~\ref{fig:harris_spectrum_growth}(b), the dominant modes display a clear
exponential phase with $\gamma/\Omega_{ci}$ in the range $0.25$--$0.35$, consistent
with linear expectations for LHDI in this regime.

\subsection{Linear evolution in shear-layer configurations}

In the velocity–shear configuration, both the lower–hybrid drift instability (LHDI) and
the Kelvin–Helmholtz instability (KHI) are linearly unstable. LHDI originates at the
density gradients on the flanks of the layer, producing short-wavelength
fluctuations with $k_\perp\rho_e\sim 1$, whereas KHI arises from the velocity shear
in the central region and favors longer wavelengths ($k_y\rho_i\sim 1$--$2$) with a
slower linear growth. Although weaker in its initial rate, KHI persists over a
longer interval in cases with reduced density contrast or enhanced velocity shear.
\begin{figure}
\centering

\begin{subfigure}{0.49\textwidth}
    \centering
    \includegraphics[width=\textwidth]{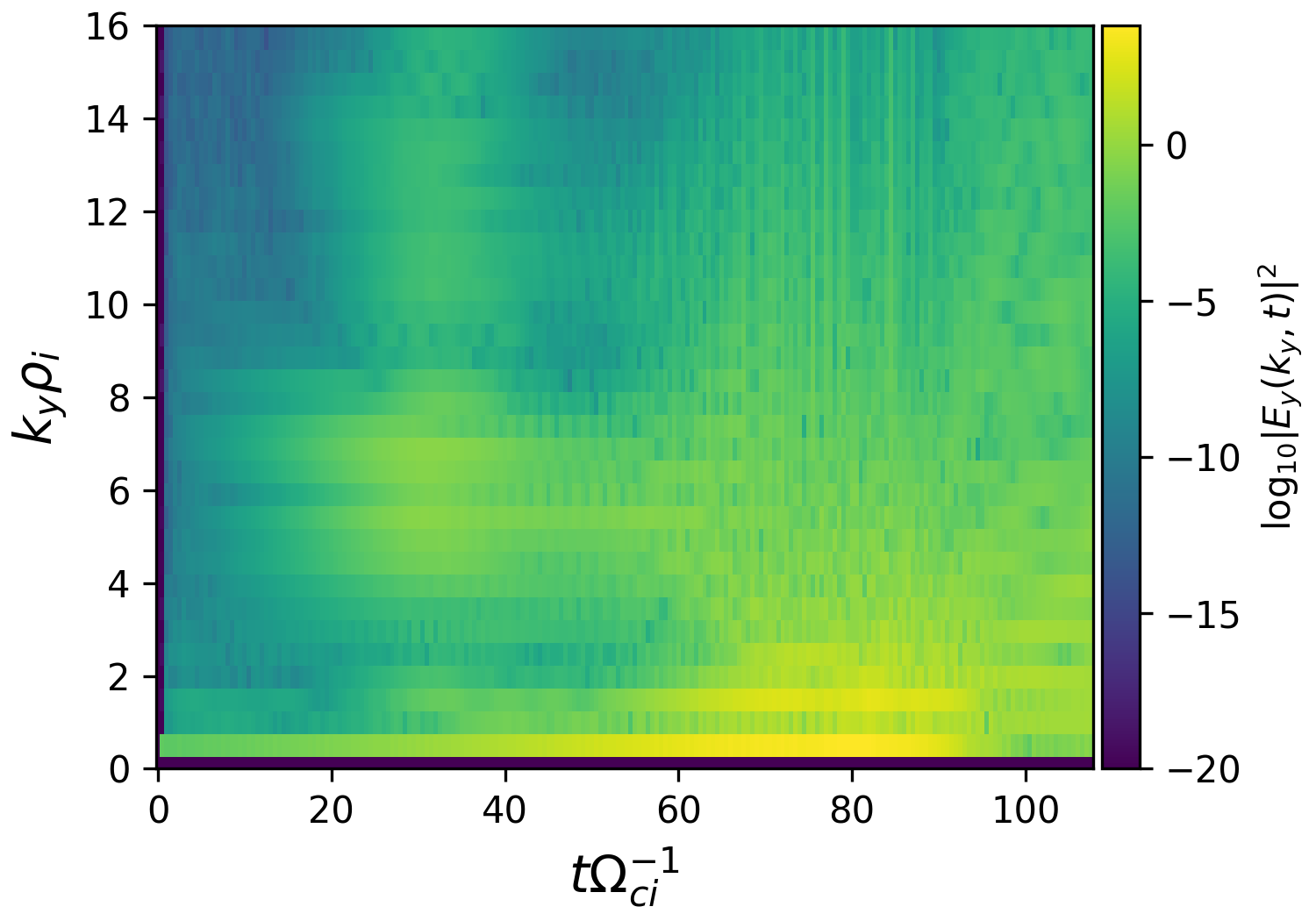}
\end{subfigure}
\hfill
\begin{subfigure}{0.49\textwidth}
    \centering
    \includegraphics[width=\textwidth]{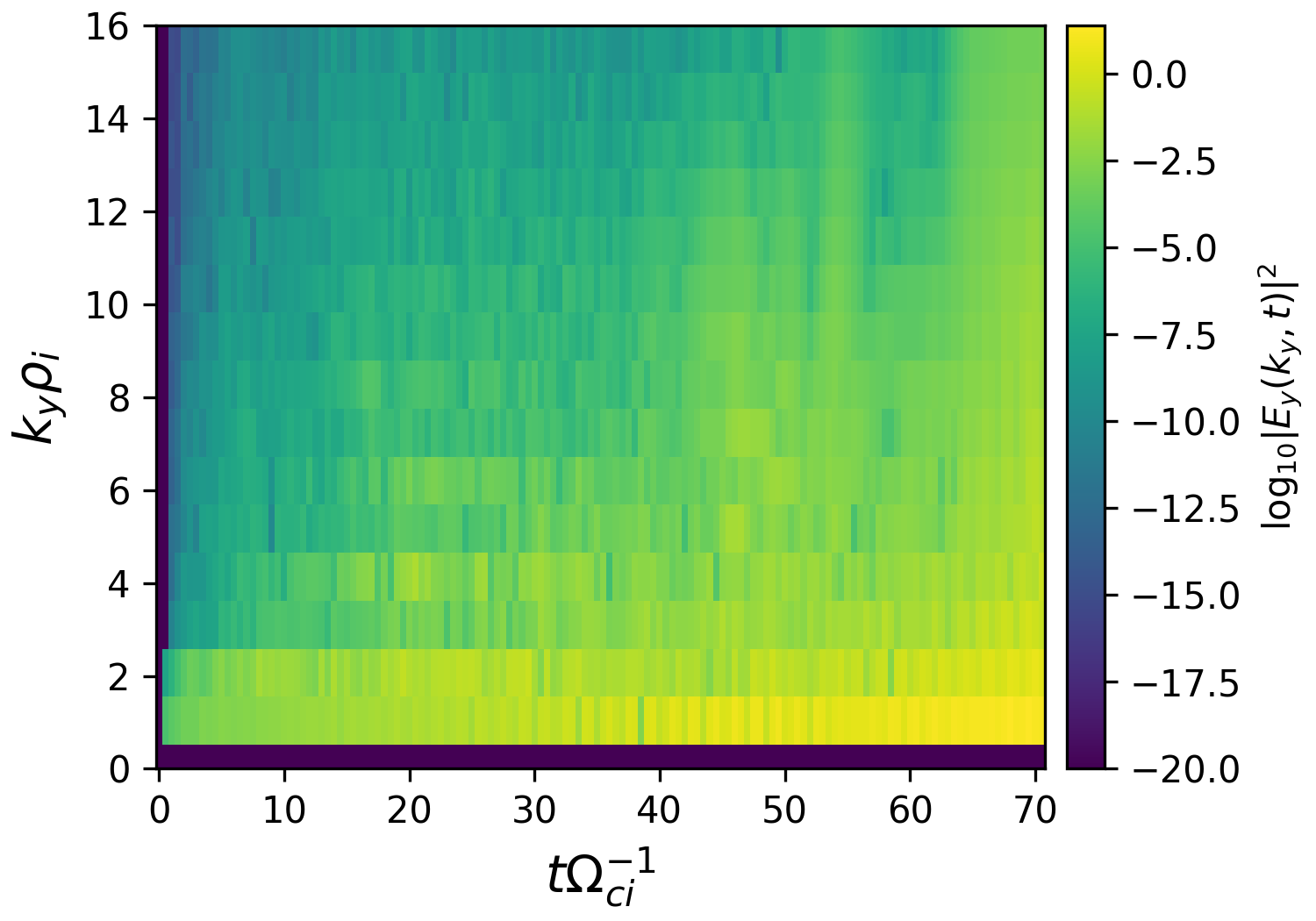}
\end{subfigure}
\hfill
\begin{subfigure}{0.49\textwidth}
    \centering
    \includegraphics[width=\textwidth]{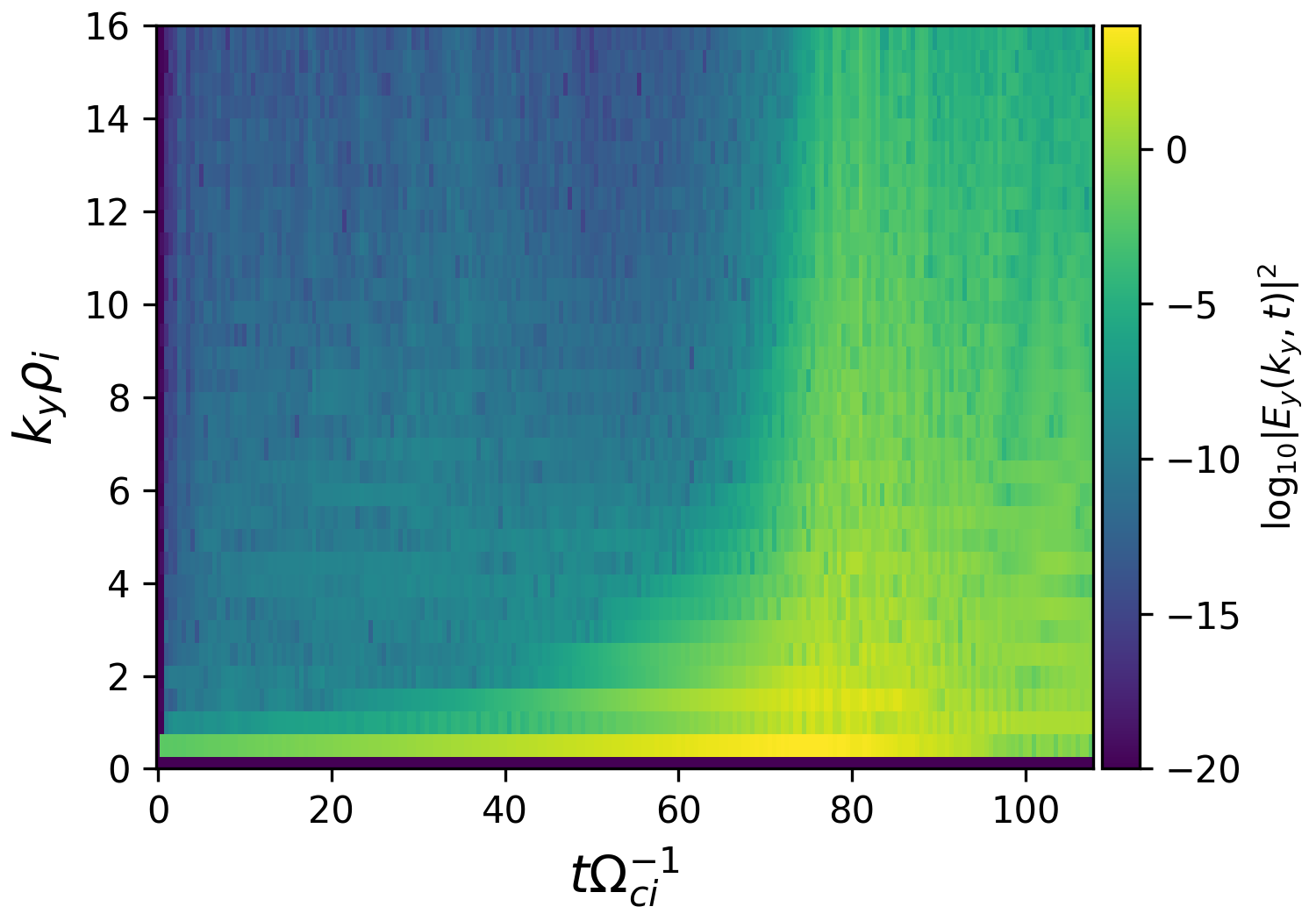}
\end{subfigure}

\caption{Electric–field spectral energy $E_{E_y}(k_y,t)$ for the three
shear–layer configurations. (a) LHDI-dominated case showing early high-$k_y$
fluctuations. (b) Mixed case where both instability families contribute. (c)
KHI-dominated case showing the gradual emergence of low-$k_y$ KH modes.}
\label{fig:shear_spectra}
\end{figure}
Three distinct regimes are observed depending on the imposed parameters $(V_0,\Delta n)$:
\begin{enumerate}
  \item \textbf{LHDI-dominated regime}:  
        LHDI grows rapidly on the layer flanks and reaches nonlinear amplitude before
        KHI becomes significant. The resulting small-scale eddies distort the shear
        profile and effectively suppress KH roll-up.

  \item \textbf{Mixed regime}:  
        LHDI and KHI grow concurrently. LHDI develops along the steepened edges of
        evolving KH vortices, while KHI reshapes the background density and velocity
        variations, modifying the local conditions for LHDI.

  \item \textbf{KHI-dominated regime}:  
        In weak-gradient or strong-shear cases, KHI emerges cleanly with minimal
        interference from LHDI. Large-scale KH vortices form and subsequently drive
        secondary fluctuations.
\end{enumerate}

Figure~\ref{fig:shear_spectra} shows the electric–field spectral energy
$E_{E_y}(k_y,t)$ for the three shear–layer configurations. The spectra clearly
distinguish the regimes: the LHDI–dominated case exhibits an early high-$k_y$
enhancement analogous to the Harris sheet; the KHI–dominated case displays a delayed
but pronounced low-$k_y$ branch; and the mixed case shows contributions from both
bands. As in the Harris configuration, linear growth rates were computed from the
temporal evolution of Fourier amplitudes, but are not plotted here; they show clear
exponential phases in each regime.
\begin{figure}
\centering

\begin{subfigure}{0.47\textwidth}
    \centering
    \includegraphics[width=\textwidth]{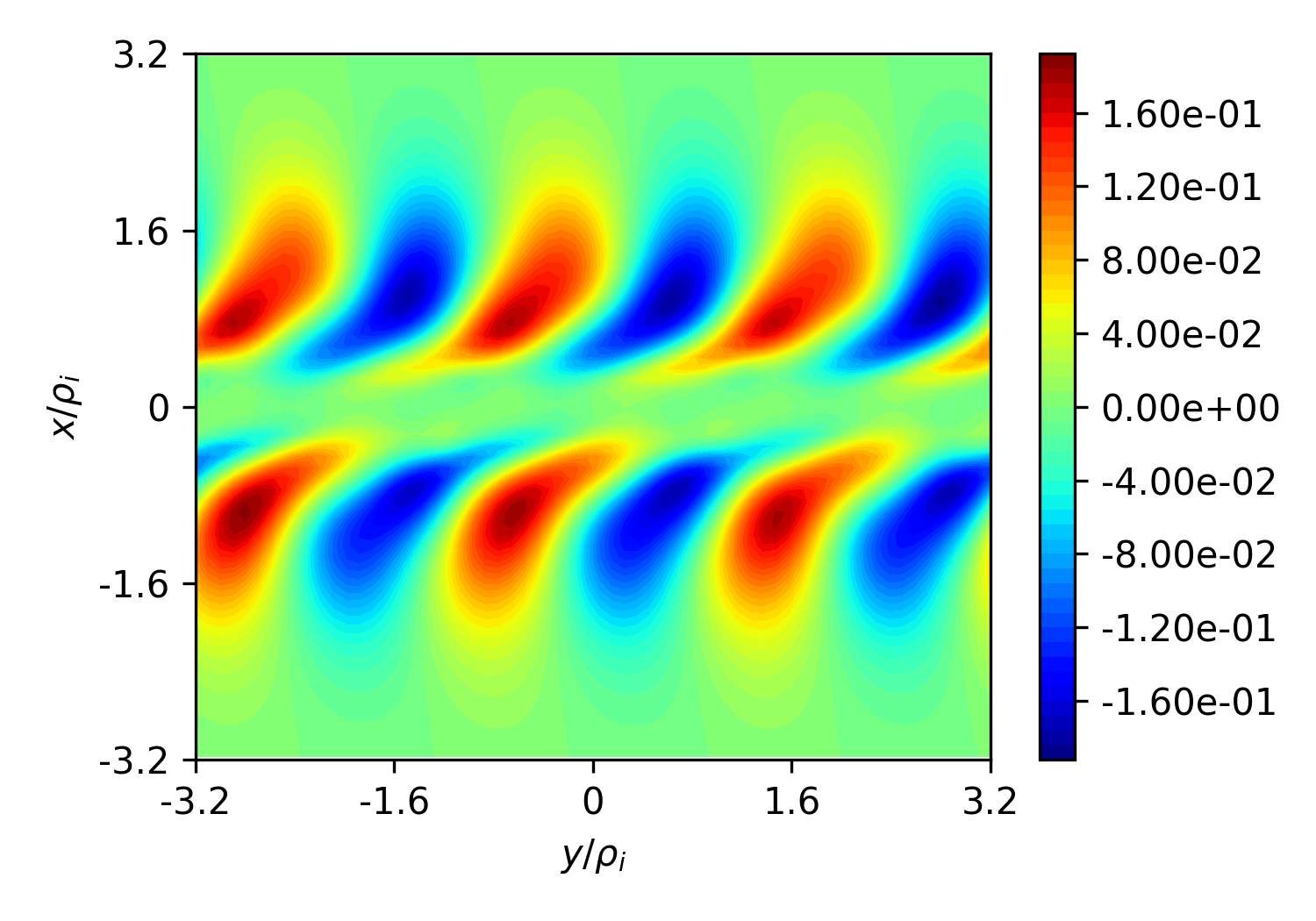}
\end{subfigure}
\hfill
\begin{subfigure}{0.47\textwidth}
    \centering
    \includegraphics[width=\textwidth]{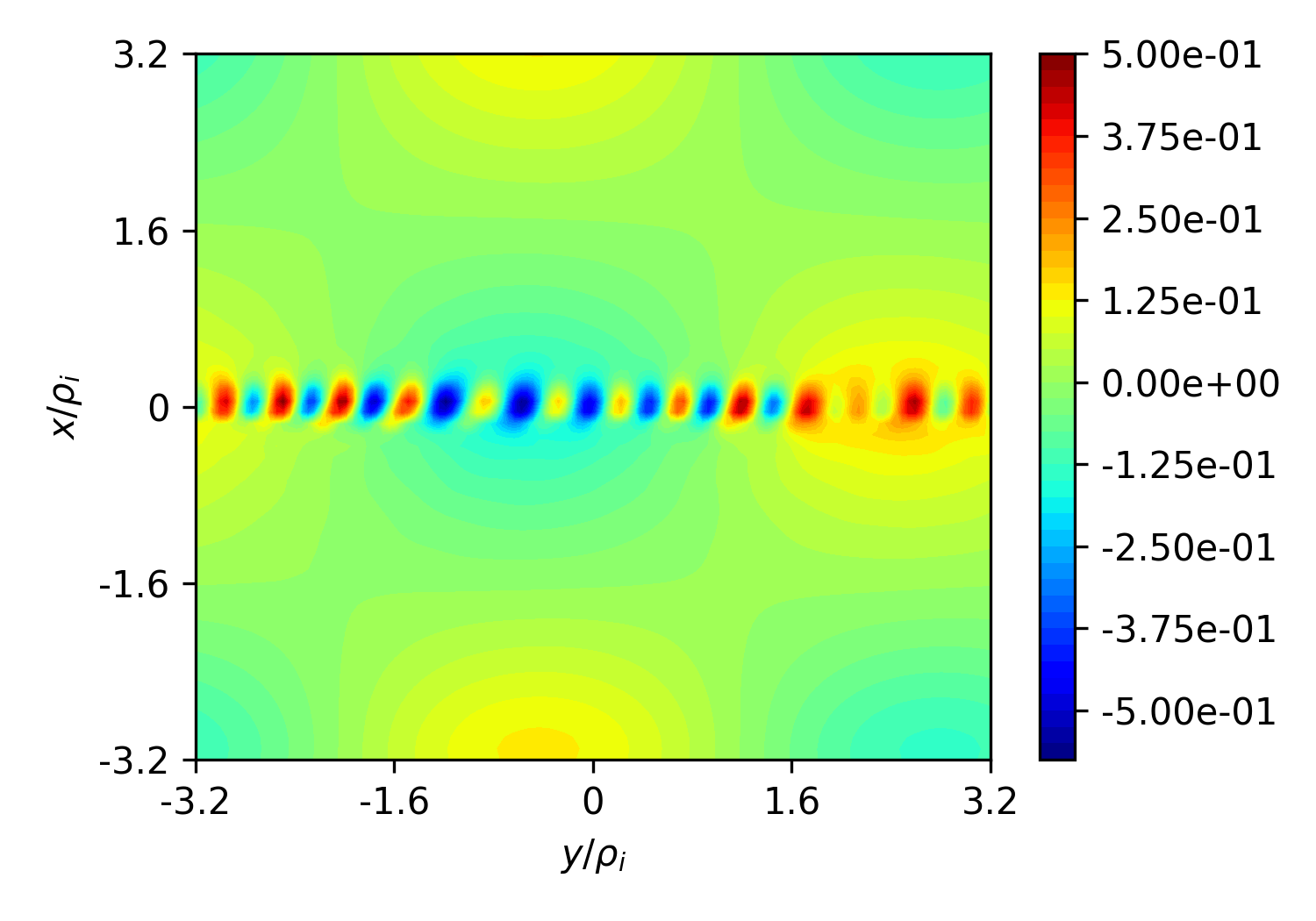}
\end{subfigure}
\hfill
\begin{subfigure}{0.47\textwidth}
    \centering
    \includegraphics[width=\textwidth]{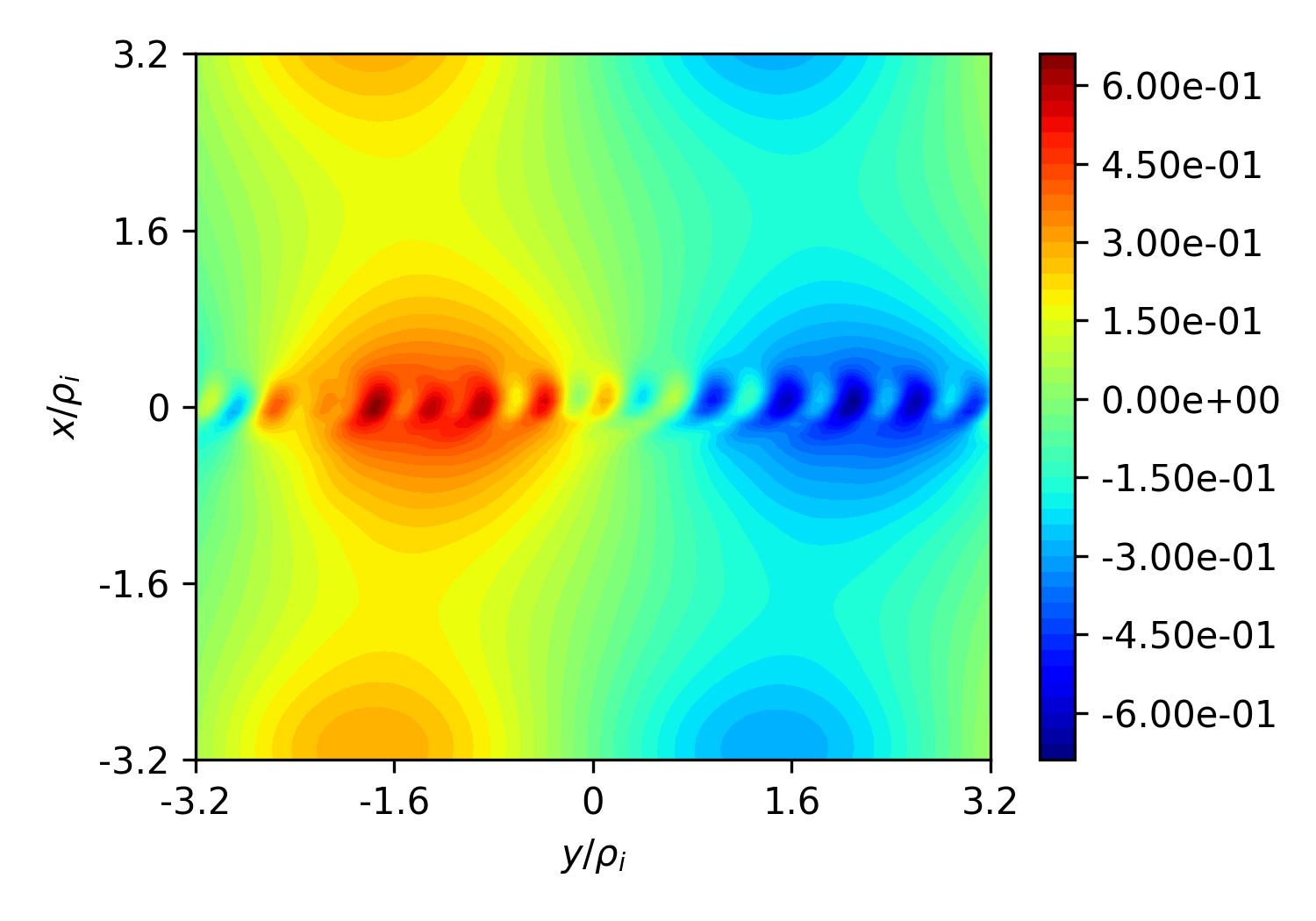}
\end{subfigure}
\hfill
\begin{subfigure}{0.47\textwidth}
    \centering
    \includegraphics[width=\textwidth]{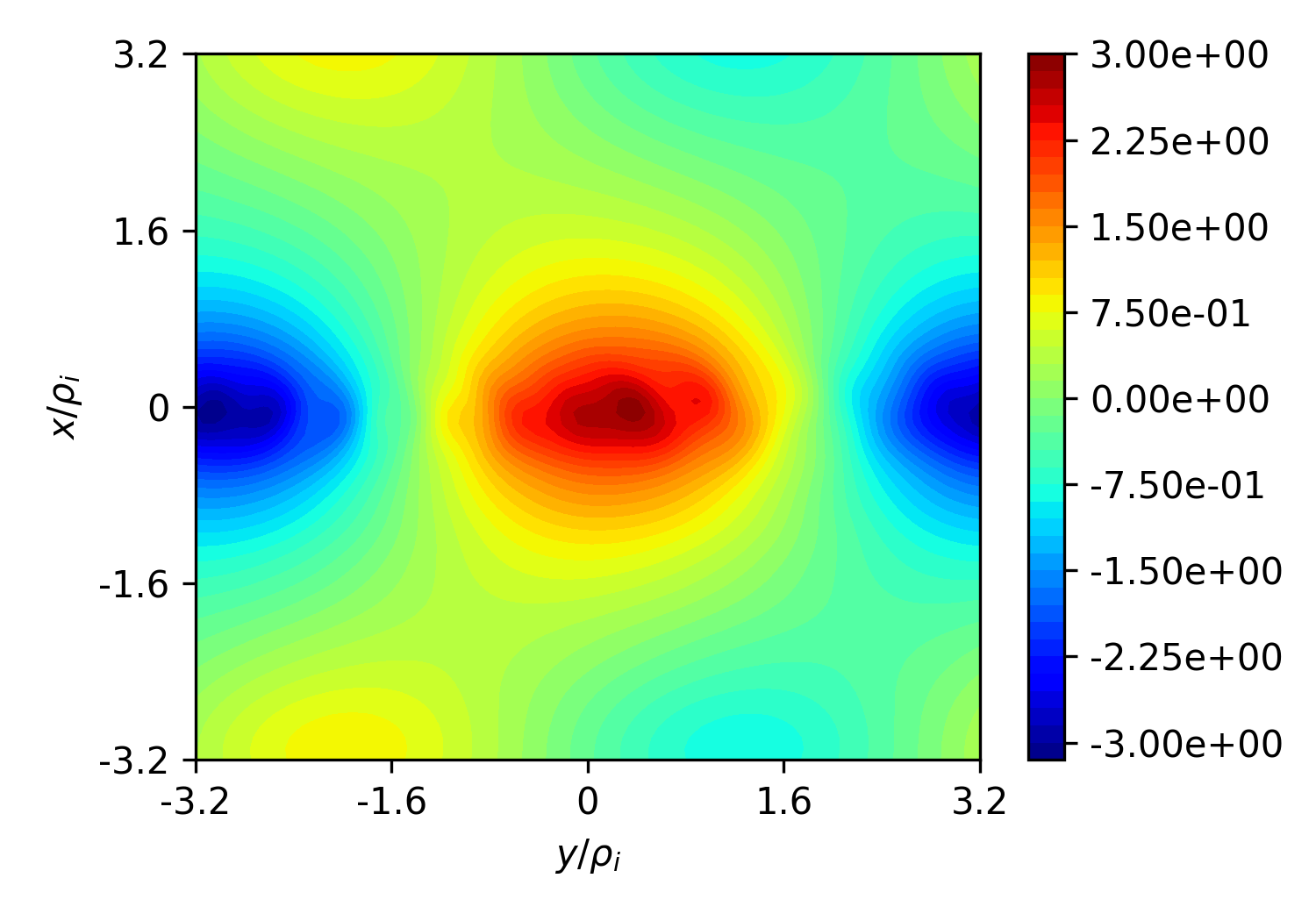}
\end{subfigure}

\caption{Two–dimensional $E_y(x,y)$ contours during the early phase 
($t \lesssim 40$-$80\,\Omega_{ci}^{-1}$) for all four configurations. 
(a) Harris sheet showing LHDI at the density–gradient edges. 
(b) Shear layer with LHDI-dominated behavior. 
(c) Mixed case with coexisting LHDI and early KH distortions. 
(d) KHI-dominated case displaying clear Kelvin–Helmholtz roll-up.}
\label{fig:Ey_contours}
\end{figure}

We further examined the evolving fluctuation structure using two–dimensional contours 
of $E_y(x,y)$ during the early phase ($t \lesssim 40$--$80\,\Omega_{ci}^{-1}$). 
Figure~\ref{fig:Ey_contours} summarizes the spatial morphology for all four 
configurations. In the Harris sheet Figure~\ref{fig:Ey_contours}(a), LHDI activity 
remains confined to the steep density gradients at the sheet edges. In the 
LHDI-dominated shear layer Figure~\ref{fig:Ey_contours}(b), similar edge–localized 
fluctuations appear, but the underlying velocity shear causes an asymmetric 
modulation of the mode amplitude. In the mixed case 
Figure~\ref{fig:Ey_contours}(c), both LHDI and early KH distortions coexist, with 
LHDI fluctuations forming along the flanks of emerging KH vortices as shown in Figure~\ref{fig:shear_plasmoids}(b). Finally, in the 
KHI-dominated configuration Figure~\ref{fig:Ey_contours}(d), large-scale KH 
roll-up forms cleanly in the central shear region while only weak LHDI signatures 
persist near the gradients.

The transition from linear to nonlinear evolution occurs earliest in the 
LHDI-dominated case, where small-scale fluctuations rapidly transfer energy to 
mesoscales through nonlinear coupling. This transition shapes the subsequent 
inverse-cascade processes and ultimately leads to plasmoid formation, as discussed 
in the next section.

\section{Nonlinear Evolution and Plasmoid Formation}
\label{sec:nonlinear}

After the linear growth stage, all four configurations enter a nonlinear regime in
which mode coupling, vortex formation, and magnetic-island generation become
prominent. The nature of this evolution depends on whether LHDI, KHI, or a
combination of both controls the early dynamics.

\subsection{Nonlinear Spectral Evolution}

\begin{figure}
\centering

\begin{subfigure}{0.48\textwidth}
    \centering
    \includegraphics[width=\textwidth]{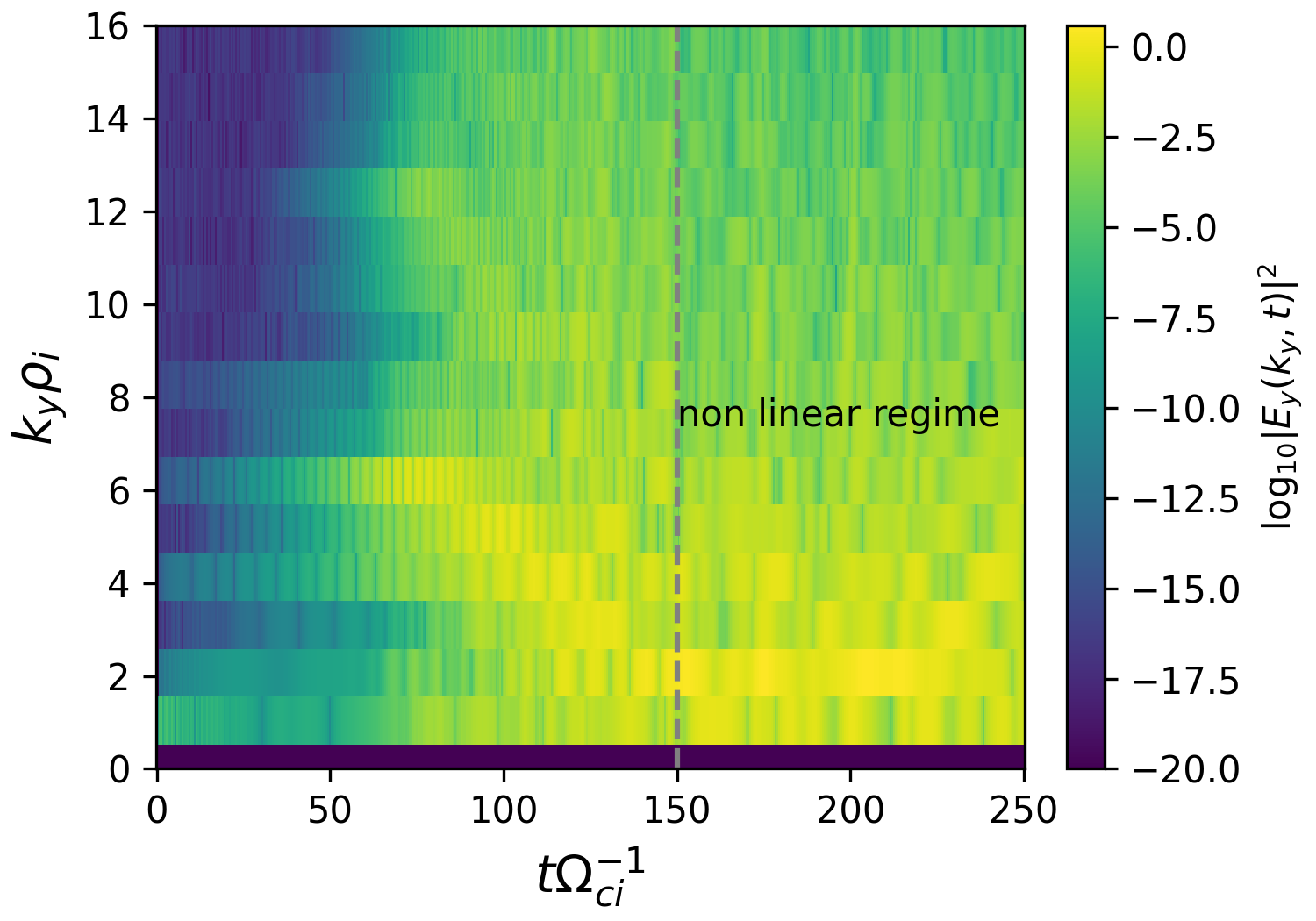}
\end{subfigure}
\hfill
\begin{subfigure}{0.48\textwidth}
    \centering
    \includegraphics[width=\textwidth]{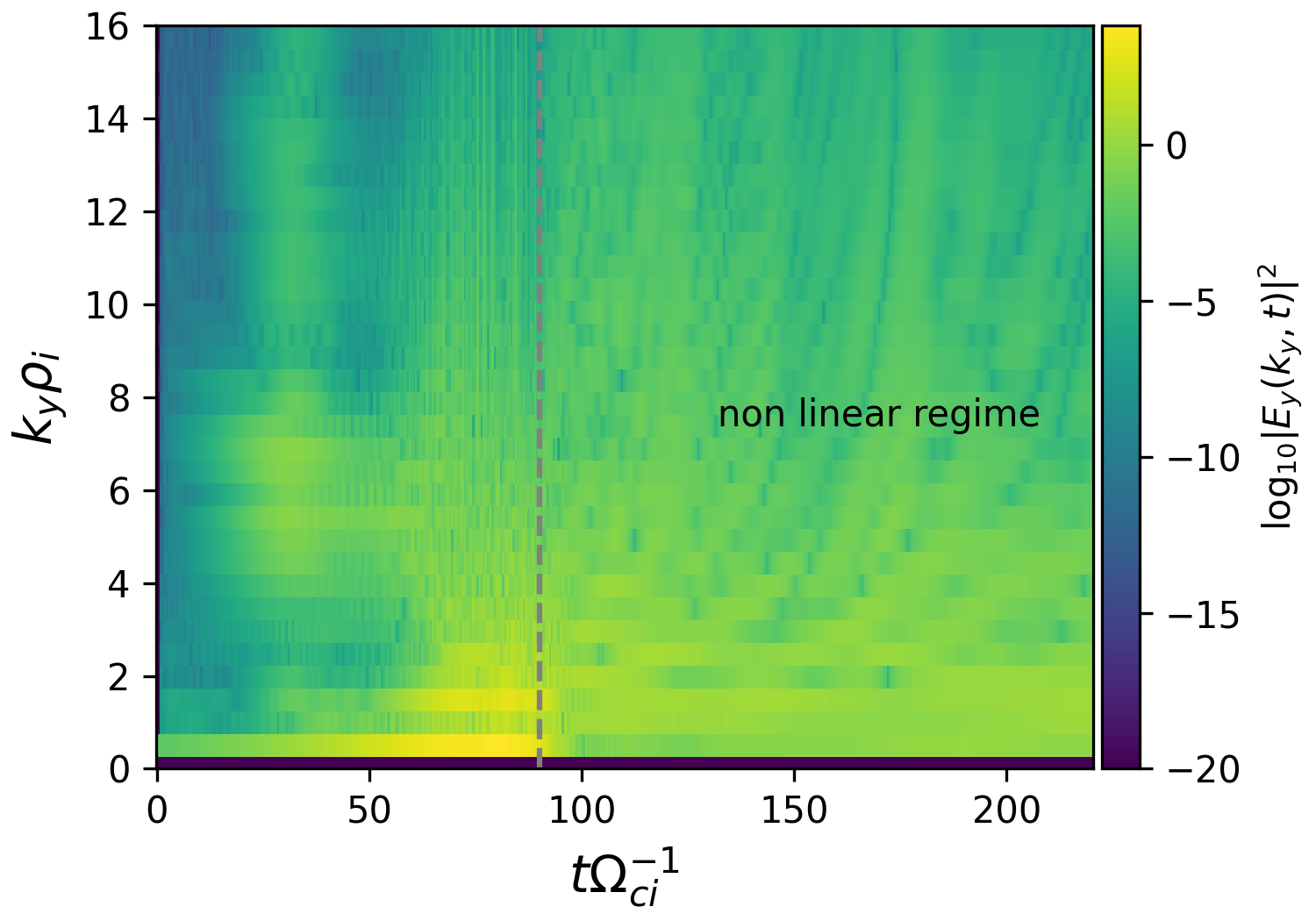}
\end{subfigure}

\vspace{0.3cm}

\begin{subfigure}{0.48\textwidth}
    \centering
    \includegraphics[width=\textwidth]{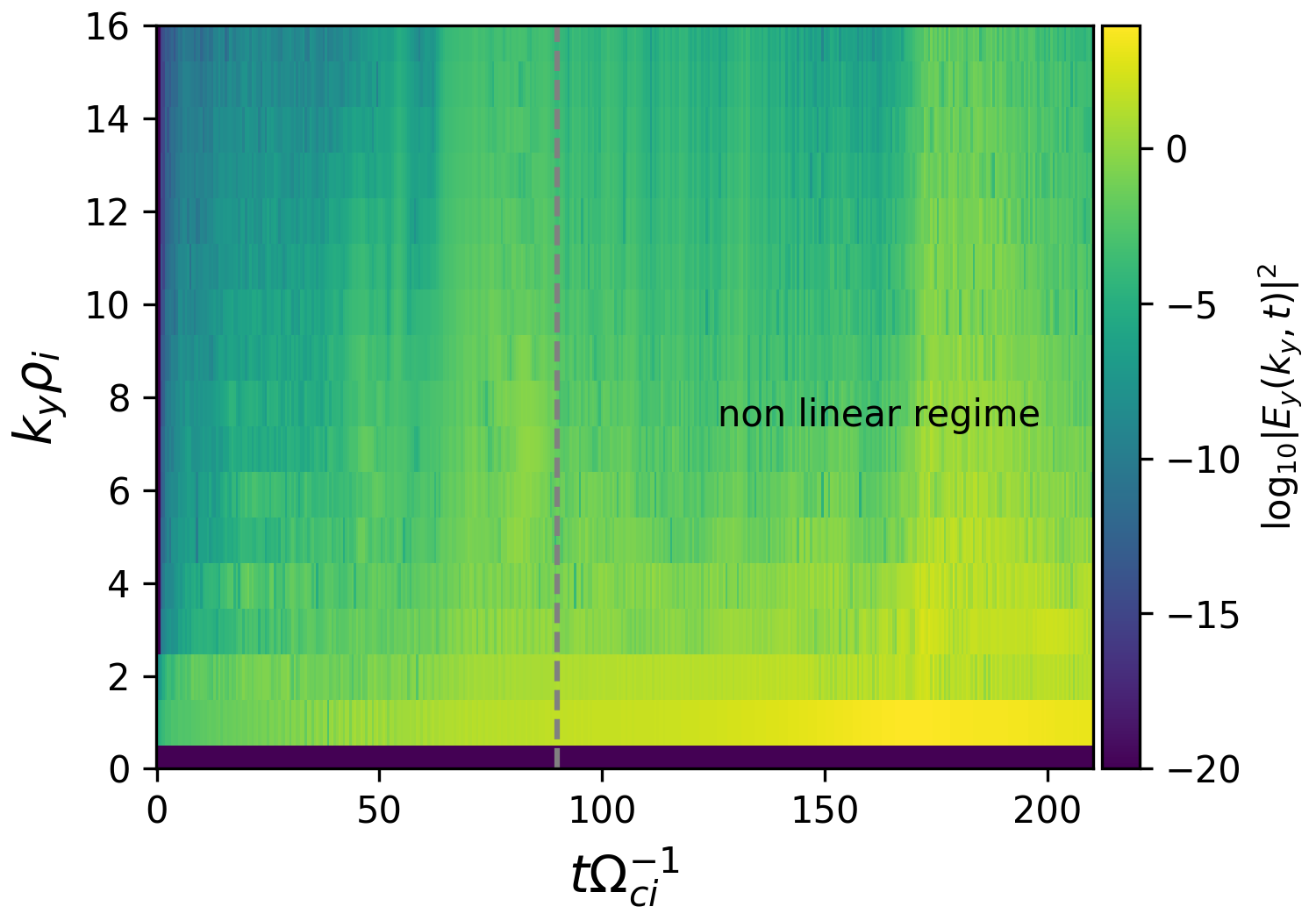}
\end{subfigure}
\hfill
\begin{subfigure}{0.48\textwidth}
    \centering
    \includegraphics[width=\textwidth]{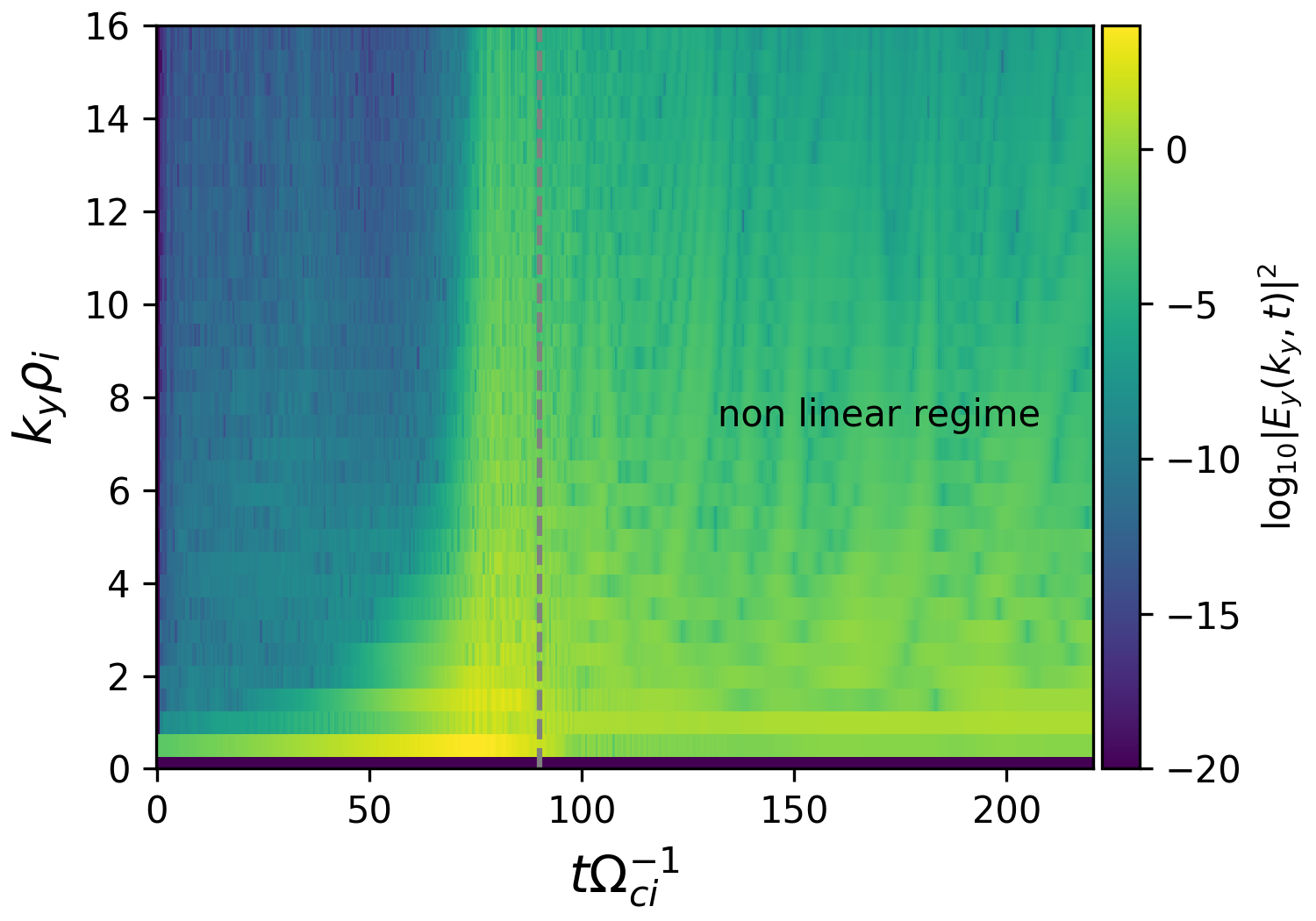}
\end{subfigure}

\caption{
Nonlinear spectral evolution of the electric field $E_{E_y}(k_y,t)$ for the four
configurations. (a) Harris sheet, (b) LHDI-dominated shear layer, 
(c) mixed regime, and (d) KHI-dominated shear layer. 
LHDI-driven cases show transfer from high to low $k_y$ during saturation, 
whereas KHI-driven cases retain dominant low-$k_y$ power until vortex breakdown.
}
\label{fig:nonlinear_spectra}
\end{figure}

The transition from linear to nonlinear behavior is accompanied by a redistribution of
spectral energy. Figure~\ref{fig:nonlinear_spectra} shows the electric--field spectral
evolution $E_{E_y}(k_y,t)$ for all four configurations across the linear–nonlinear
boundary. In the Harris sheet and in the LHDI-dominated shear layer, energy initially
concentrated at high $k_y$ progressively transfers toward lower $k_y$ as nonlinear
interactions broaden the spectrum. In contrast, the KHI-dominated configuration shows
sustained low-$k_y$ power associated with KH roll-up, with small-scale fluctuations
appearing only after vortex disruption. The mixed case exhibits both behaviors:
LHDI-driven high-$k_y$ activity along the layer flanks and KHI-driven low-$k_y$
structure in the core.
\subsection{Harris Sheet: LHDI-Driven Nonlinear Evolution}

\begin{figure}
\centering
\includegraphics[width=0.5\textwidth]{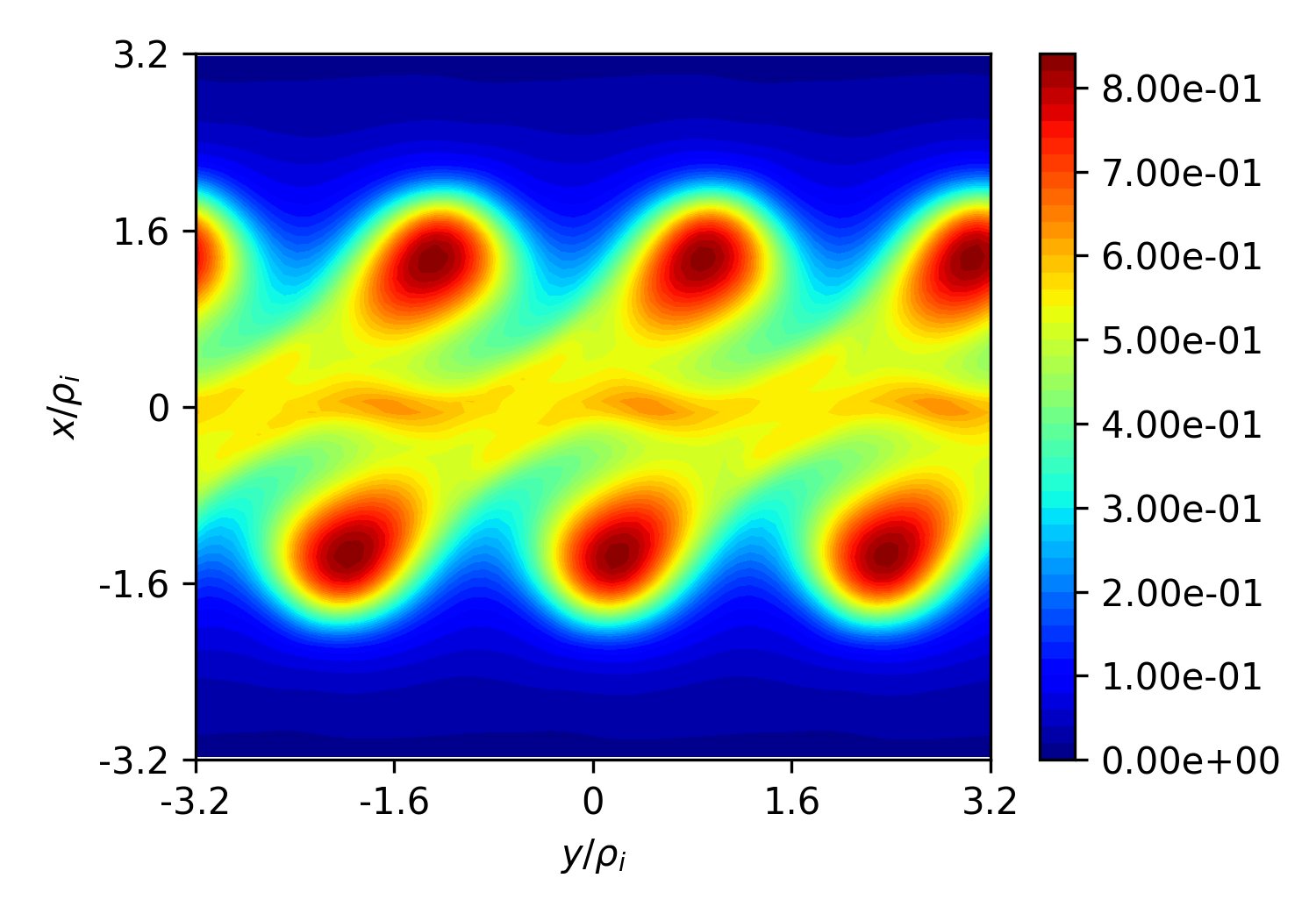}
\caption{Nonlinear evolution in the Harris sheet. $J_z$ highlighting LHDI-driven filamentation and showing closed magnetic islands emerging without a tearing-mode seed.}
\label{fig:harris_plasmoids}
\end{figure}

In the Harris configuration, LHDI saturates rapidly and generates a turbulent
state along the sheet flanks. Small-scale current filaments and electric-field
vortices merge through nonlinear interaction, forming progressively larger
structures via an inverse cascade. Representative snapshot of $J_z(x,y)$ (Figure~\ref{fig:harris_plasmoids}) show magnetic islands forming along
the sheet edges where LHDI was strongest.

\subsection{Shear Layer: Three Nonlinear Regimes}

\begin{figure}
\centering
\begin{subfigure}{0.49\textwidth}
\includegraphics[width=\textwidth]{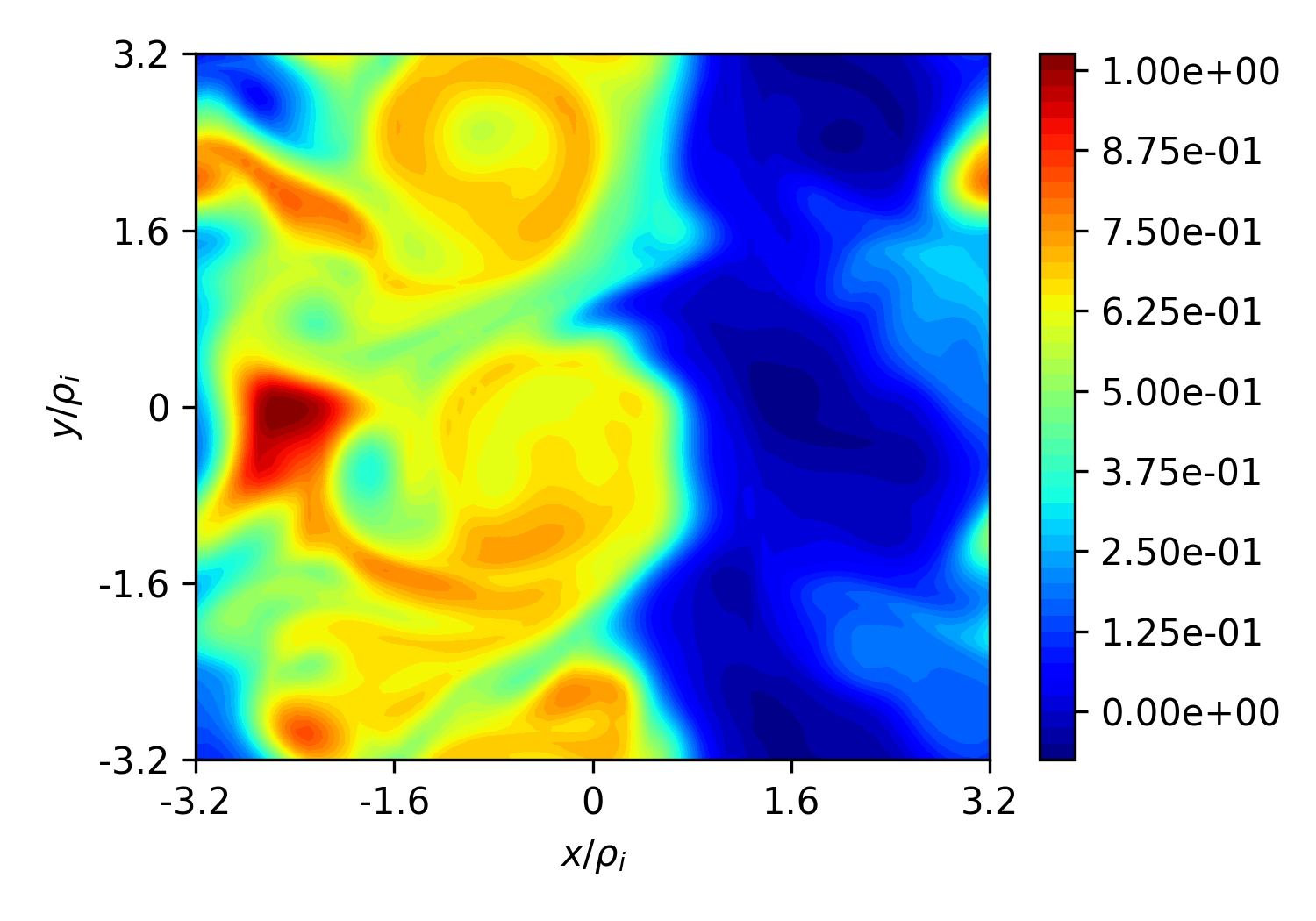}
\end{subfigure}
\hfill
\begin{subfigure}{0.49\textwidth}
\includegraphics[width=\textwidth]{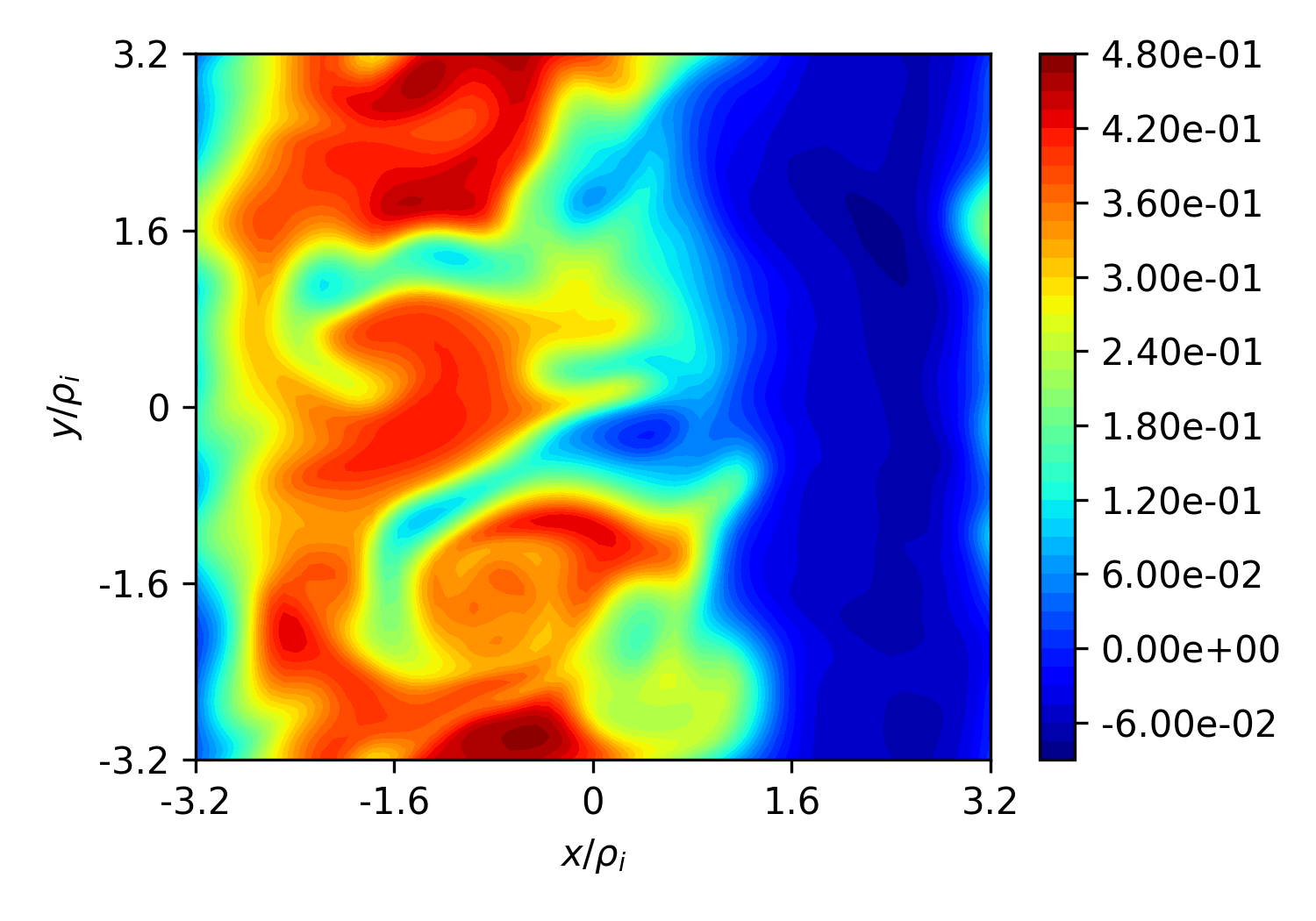}
\end{subfigure}
\hfill
\begin{subfigure}{0.49\textwidth}
\includegraphics[width=\textwidth]{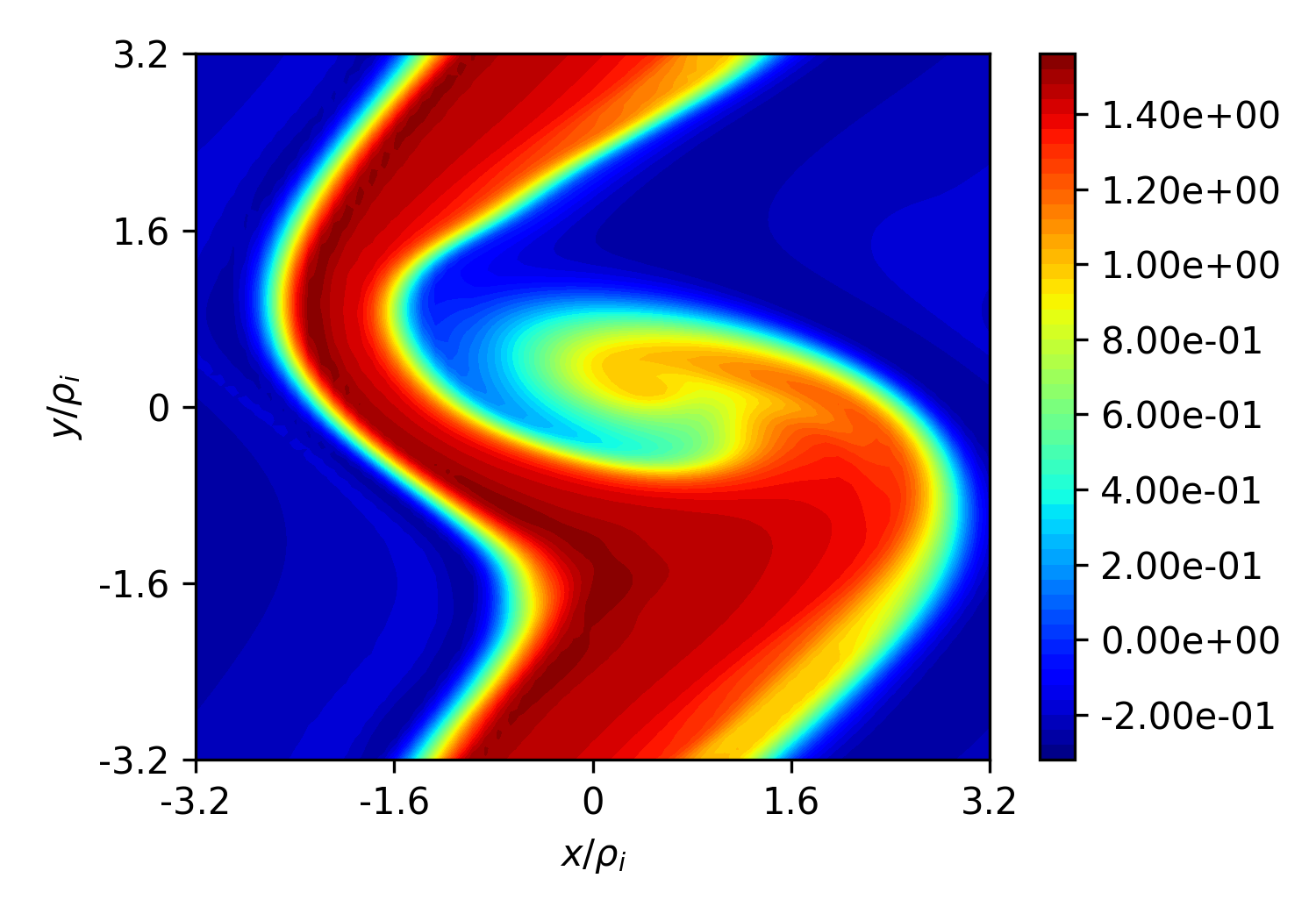}
\end{subfigure}
\caption{Nonlinear current-density structures $J_z(x,y)$ in the three shear–layer 
configurations: (a) LHDI-dominated, (b) mixed regime, and (c) KHI-dominated. 
LHDI-driven cases show edge-localized current filaments, the mixed regime exhibits
both LHDI and KH deformation, and the KHI-dominated case forms large vortex-centered
current sheets that reconnect and generate plasmoids.}
\label{fig:shear_plasmoids}
\end{figure}

The nonlinear evolution of the shear layer depends sensitively on the parameters
$(V_0,\Delta n)$ that determine the relative strength of LHDI and KHI. Representative
snapshots of the out–of–plane current density $J_z(x,y)$ for the three shear–layer
configurations are shown in Figure~\ref{fig:shear_plasmoids}. These panels highlight
the different pathways through which magnetic structure and plasmoid formation emerge.

\begin{figure}
\centering

\begin{subfigure}[t]{0.49\textwidth}
    \centering
    \includegraphics[width=\textwidth]{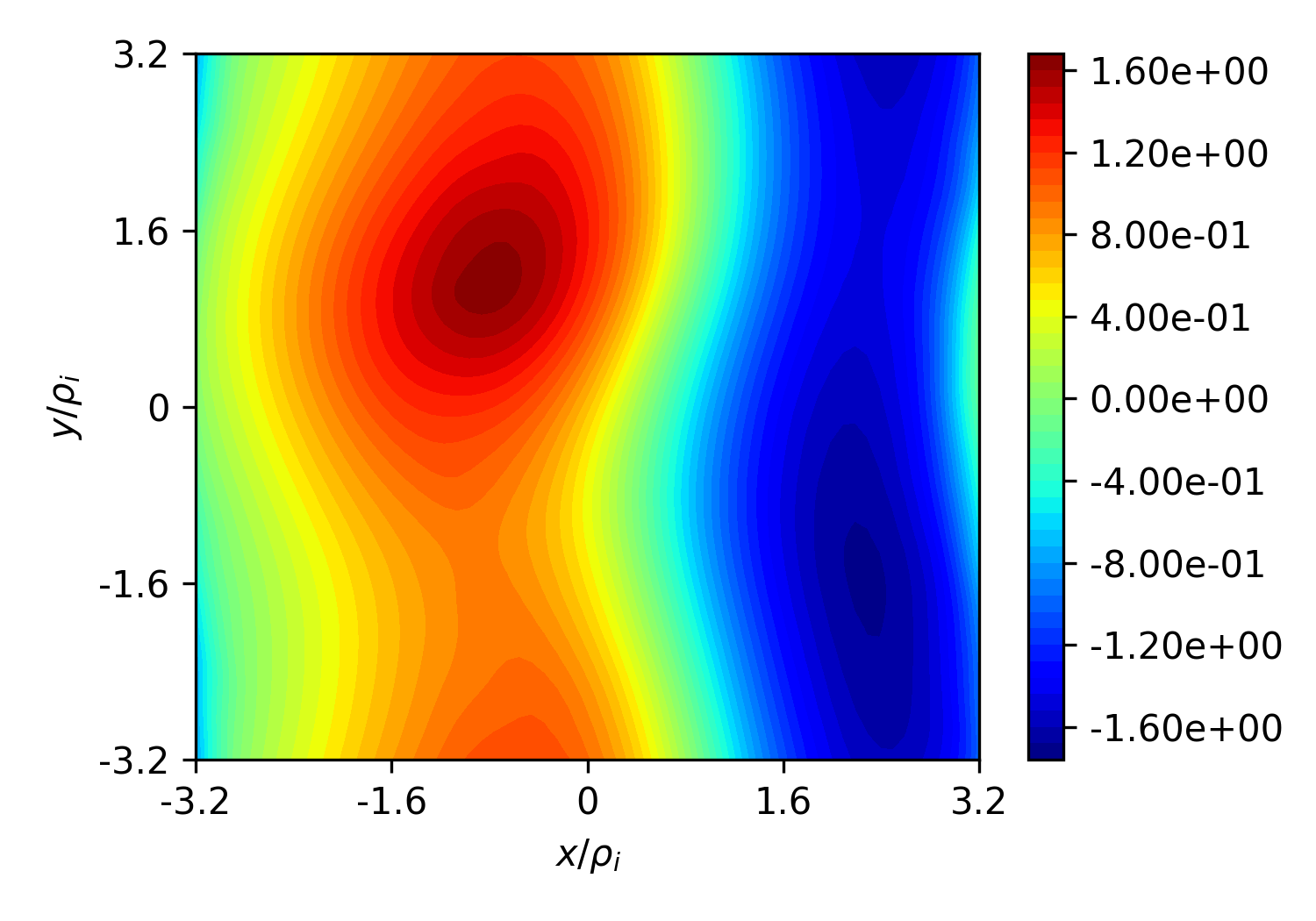}
\end{subfigure}
\hfill
\begin{subfigure}[t]{0.49\textwidth}
    \centering
    \includegraphics[width=\textwidth]{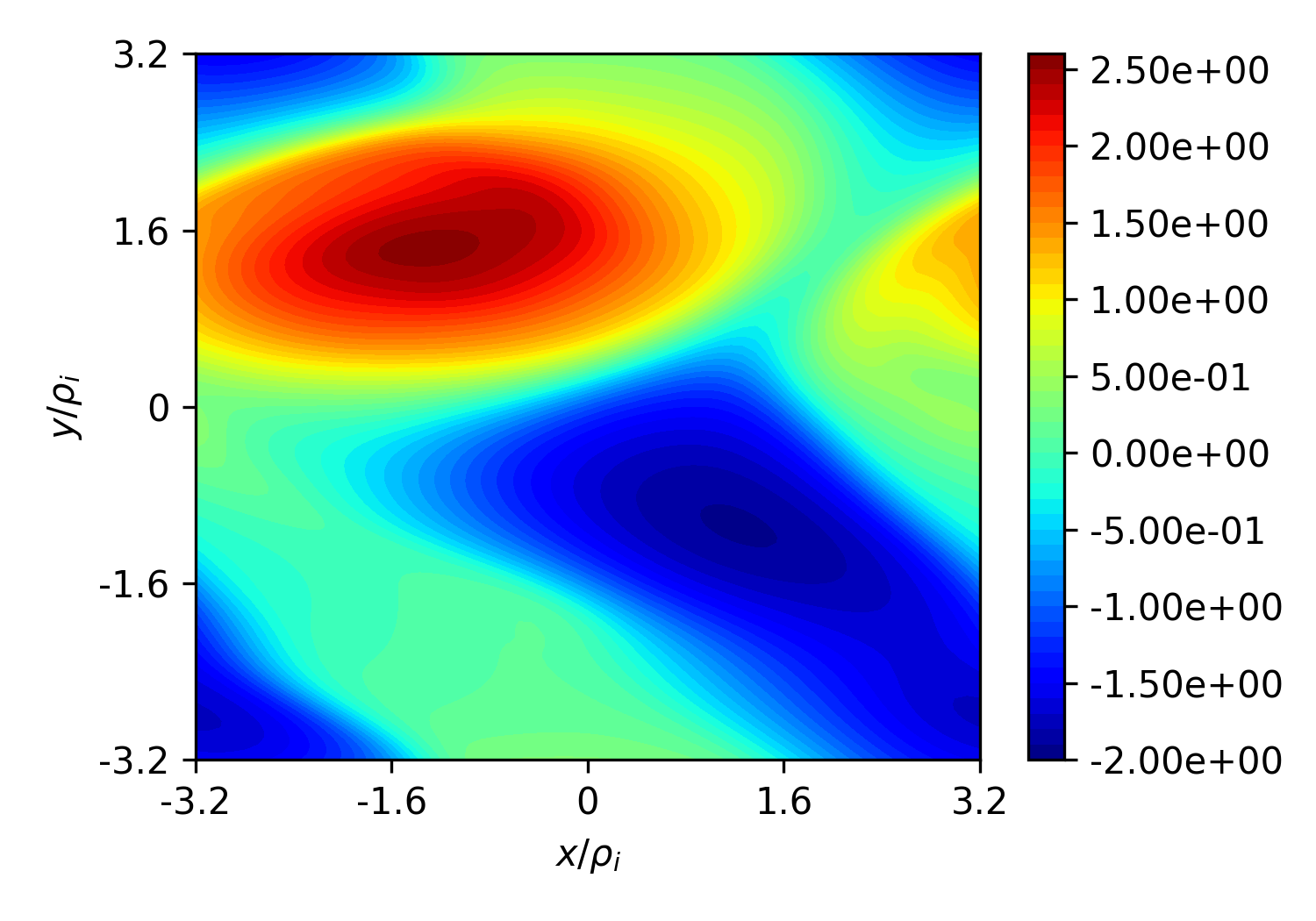}
\end{subfigure}

\caption{Internal reconnection and plasmoid formation in the KHI-dominated shear 
layer, illustrated using the parallel vector potential $A_z(x,y)$. The left panel 
shows the initial field reversal in a rolled-up KH vortex, while the right panel 
shows the subsequent formation of a coherent magnetic island produced by secondary 
reconnection.}
\label{fig:kh_plasmoids}
\end{figure}

In the LHDI-dominated case Figure~\ref{fig:shear_plasmoids}(a), fluctuations remain
confined to the density gradients at the flanks of the layer, where they broaden the
shear profile and delay the onset of KHI. In the mixed regime 
Figure~\ref{fig:shear_plasmoids}(b), LHDI forms along the steepened edges of early KH
vortices while KHI simultaneously distorts the core region, creating favorable
conditions for secondary LHDI. In the KHI-dominated configuration 
Figure~\ref{fig:shear_plasmoids}(c), large KH vortices roll up cleanly in the center of the layer. As they
mature, these vortices stretch the in–plane magnetic field and generate
secondary current layers along their flanks. These secondary layers
reconnect and produce plasmoids inside the vortex cores, as illustrated
in Figure~\ref{fig:kh_plasmoids}. The resulting magnetic islands are larger
and more coherent than those found in the LHDI-driven cases.

\subsection{Plasmoid Statistics}
\begin{figure}
\centering
\includegraphics[width=0.75\textwidth]{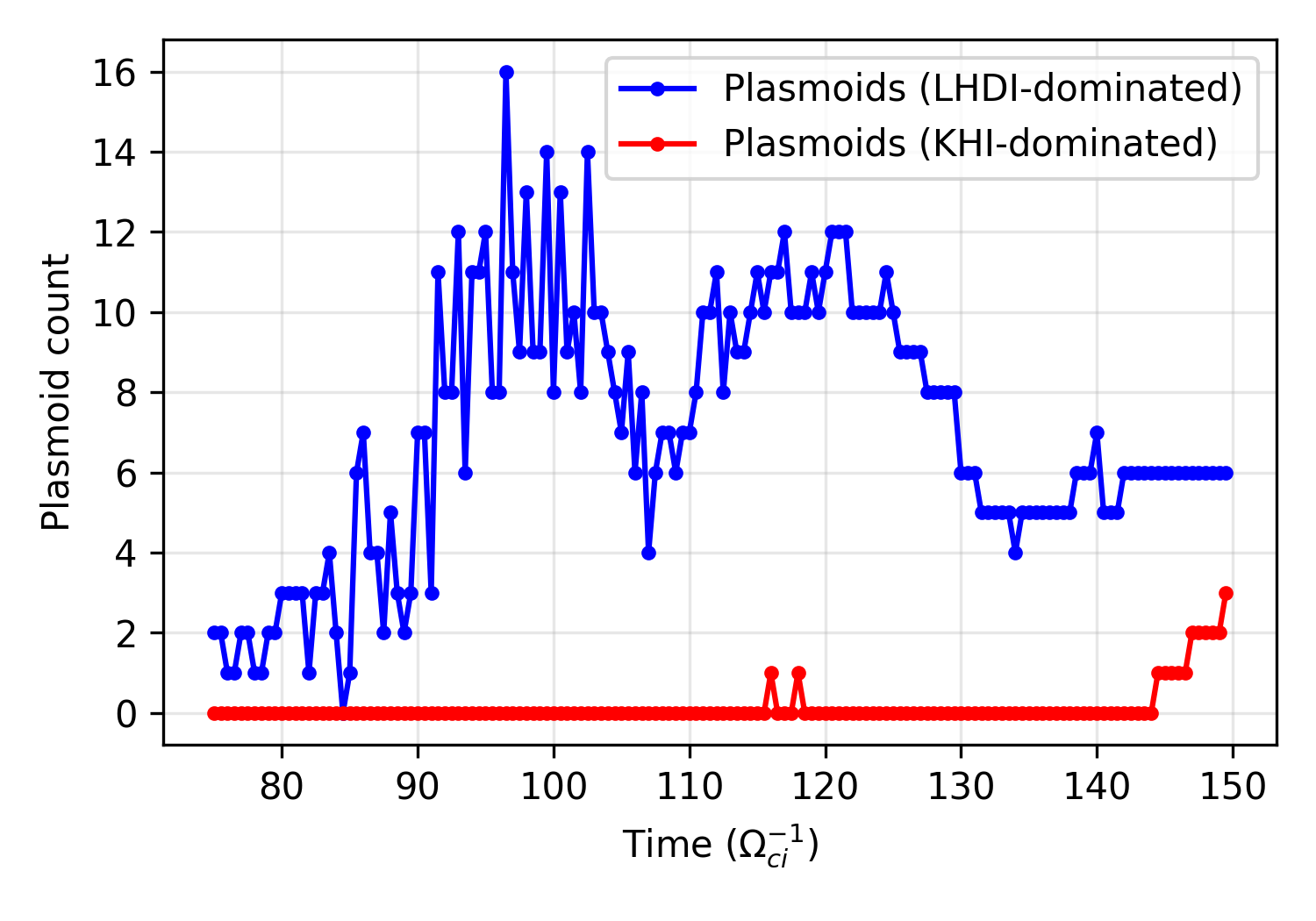}
\caption{Temporal evolution of plasmoid count in nonlinear phase. Blue:
LHDI-dominated shear layer; red: KHI-dominated shear layer.}
\label{fig:plasmoid_count}
\end{figure}

Plasmoids are identified using topological analysis of $A_z$ following
Ref.~\citep{Huang2010}. LHDI-driven cases produce earlier and more numerous
plasmoids, while KHI-driven plasmoids appear later and are larger. Time series of
plasmoid number in nonlinear phase are shown in 
Figure~\ref{fig:plasmoid_count}.
A typical LHDI dominated shear layer case yields $\sim$14-16 plasmoids by $t = 100\,\Omega_{ci}^{-1}$, while KHI dominated shear layer runs generate $\sim$2–4 plasmoids by $t = 150\,\Omega_{ci}^{-1}$. Time series of plasmoid number and reconnection electric field are shown in Figure~\ref{fig:plasmoid_count}. 

These results highlight the multiple pathways to plasmoid generation in collisionless systems and emphasize the role of microinstability-driven turbulence in initiating reconnection and shaping current sheet evolution.

\section{Discussion}
\label{sec:discussion}

The simulations presented here illustrate how the nonlinear interplay between LHDI and KHI governs plasmoid formation pathways in low-$\beta_e$, collisionless current sheets and shear layers. This section discusses our results in the broader context of existing theory, simulations, and observational evidence, and explores the physics implications of the inverse cascade mechanism, instability competition, and turbulence-driven reconnection.

\subsection{Comparison with Previous Studies}

Our findings support and extend the results of~\citet{Dargent2019}, who first reported the suppression of KHI by LHDI-driven turbulence in fully kinetic PIC simulations of the Mercury magnetopause. Using a hybrid-gyrokinetic model, we confirm that steep density gradients can cause LHDI to dominate the early dynamics, producing meso-scale structures that preempt the development of KHI.

In contrast to PIC approaches, our hybrid code allows for larger domain sizes and long-time integration, making it feasible to observe inverse cascade development, plasmoid formation, and secondary reconnection beyond the initial growth phase. The emergence of plasmoids via nonlinear LHDI activity, in the absence of tearing-mode seed perturbations, reveals an alternative mechanism for flux rope formation in environments where classical reconnection is not initially favored.

Our results also relate to two-fluid and Hall-MHD studies (e.g.,~\citet{Nakamura2020};~\citet{Roytershteyn2012}) which reported the co-existence of LHDI with tearing or kink instabilities in magnetotail current sheets. However, those works typically did not resolve the full nonlinear transition of LHDI into plasmoid-producing structures via inverse cascade. Here, we directly show that LHDI can restructure the current sheet to 
facilitate reconnection by preconditioning the current sheet, and in some cases, trigger localized onset where strong LHDI-driven turbulence forms.

The KH vortex-induced reconnection we observe in the shear layer setup agrees with findings from ~\citet{Nakamura2006},~\citet{Zhang2022},~\citet{Faganello2017}, who demonstrated that KHI can roll up and trap magnetic field lines, resulting in secondary reconnection sites within vortices. Our hybrid kinetic--gyrokinetic model reproduces this effect and adds to it by capturing the concurrent development of LHDI at vortex edges, highlighting the dynamical feedback between these instabilities.

\subsection{Inverse Cascade and Suppression of KHI}

A central feature of our LHDI-dominated runs is the presence of an inverse cascade: small-scale electric and magnetic fluctuations coalesce into larger-scale eddies that mimic KHI structures in size but arise from kinetic processes~\citep{Karimabadi2013}. This behavior was also observed by~\citet{Dargent2019}, but our study extends it by quantifying its role in plasmoid formation and by examining how it suppresses shear-driven modes.

When LHDI saturates and cascades upward in scale, it redistributes vorticity and magnetic tension in such a way that the original shear flow profile is no longer sharp. The effective reduction in velocity shear weakens the free energy available for KHI, leading to its suppression. We find this is especially effective in low-$\beta_e$ plasmas, where LHDI growth is fastest and the resulting turbulence remains electrostatic for longer.

This effect can be seen as a type of kinetic pre-conditioning: the LHDI generates structures that "occupy" the phase space and flow geometry where KHI would otherwise grow, and in doing so, modifies the evolution path of the system. 
Such coupling between kinetic microturbulence and fluid-scale instability exemplifies the cross-scale interactions that characterize multiscale reconnection dynamics.

\subsection{Relevance to Space and Astrophysical Plasmas}

In the solar wind, magnetopause, and magnetotail environments, current sheets often feature both shear flows and density gradients. Depending on the gradient scale lengths and local plasma parameters, either LHDI or KHI (or both) may be active. Observations from MMS, THEMIS, and Cluster missions have frequently reported signatures of lower-hybrid waves near reconnection sites, as well as Kelvin–Helmholtz vortices at magnetopause flanks~\citep{Hasegawa2004,Norgren2012,Wilder2019,Tang2013,Hwang2011}.

Our results suggest that LHDI may play a more central role in structure formation than previously assumed, particularly in the dissipation range. By seeding or inhibiting KHI, LHDI could determine whether large-scale eddies form or whether the system evolves directly into turbulence. This has implications for how we interpret observed boundary layer fluctuations and flux rope populations in satellite data.

Furthermore, LHDI-driven plasmoid formation provides an alternative pathway for magnetic island generation in reconnection environments lacking strong initial perturbations or dominant tearing modes. This could be important in understanding flux rope onset in weak guide field or asymmetric reconnection regions, where traditional models fall short.

\subsection{Parameter Dependencies and Scaling}

The present study does not include a parameter scan for KHI; instead, the three 
shear–layer configurations were chosen to isolate LHDI–dominated, mixed, and 
KHI–dominated regimes. The parameter dependence of LHDI itself was examined in detail 
in our previous work~\citep{Thatikonda2025}, where its sensitivity to mass 
ratio, temperature ratio, plasma beta, and density–gradient scale length was 
quantified. Here we summarize the main qualitative trends relevant to the combined 
LHDI–KHI problem:

\begin{itemize}
  \item \textbf{Mass ratio}: Increasing $m_i/m_e$ enhances the separation between 
  electron and ion kinetic scales and generally strengthens LHDI. Realistic mass 
  ratios allow the LHDI peak near $k_\perp\rho_e\!\sim\!1$ to be well resolved~\citep{Daughton2003}.
  
  \item \textbf{Plasma beta}: Lower $\beta_e$ enhances the electrostatic character 
  of LHDI and increases its growth rate, while higher $\beta_e$ reduces LHDI drive and 
  allows KHI to persist over longer intervals.
  
  \item \textbf{Temperature ratio}: Colder electrons (larger $T_i/T_e$) increase the diamagnetic-drift contrast, thereby enhancing the LHDI growth, consistent with linear theory.
  
 \item \textbf{Gradient scale lengths:} Steeper density gradients favor LHDI, whereas stronger velocity--shear amplitudes (larger $V_{0}$) promote KHI. The relative strength of $\nabla n$ and $V_{0}$ therefore determines whether the layer responds predominantly with LHDI, KHI, or a mixed mode.

\end{itemize}
\begin{figure}
\centering
\includegraphics[width=0.65\textwidth]{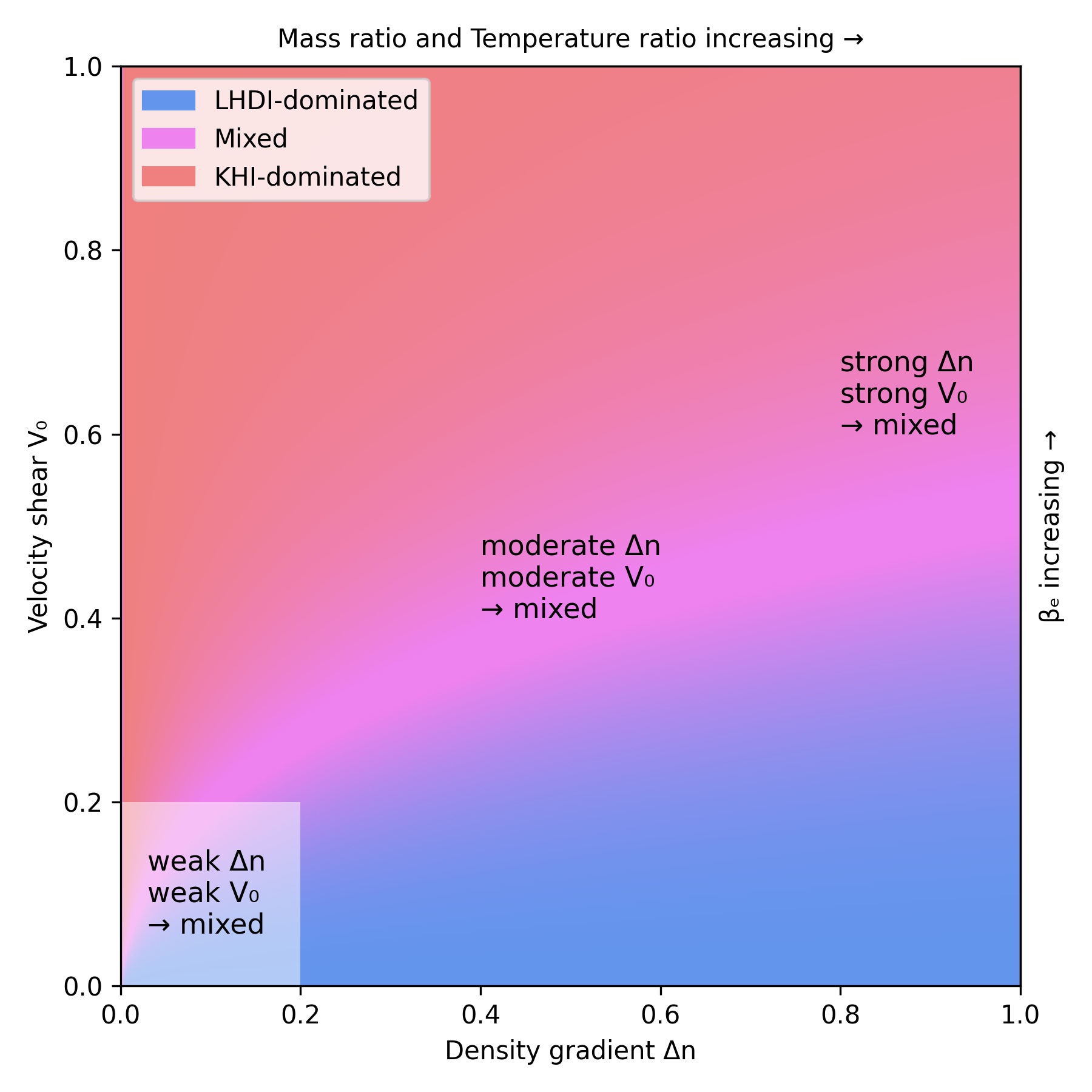}
\caption{
Conceptual scaling diagram summarizing the physical parameter trends that govern 
LHDI--KHI competition. Steep density gradients, large mass ratio, and low plasma 
$\beta_e$ promote LHDI, whereas strong or broad velocity shear and high $\beta_e$ 
favor KHI. Mixed regimes arise when the two drives are comparable or simultaneously 
weak.}
\label{fig:scaling_diagram}
\end{figure}
These parameter dependencies can be concisely represented in a schematic scaling 
diagram that highlights how the density--gradient strength and velocity--shear 
amplitude compete to favor either LHDI or KHI. In contrast to the run--selection 
map shown earlier in Figure~\ref{fig:V0_deltan_map}, the illustration in 
Figure~\ref{fig:scaling_diagram} emphasizes the underlying physical tendencies: 
steeper density gradients, larger mass ratio, and colder electrons enhance LHDI. 
However, at sufficiently large mass ratio or high $T_i/T_e$, the LHDI drive begins 
to saturate because the electron-scale response weakens relative to the ion-scale 
shear dynamics. In this regime, KHI is able to dominate, as stronger or broader 
velocity shear and higher plasma $\beta_e$ more effectively sustain shear-driven 
distortions than the diminishing LHDI drive. The mixed regime occupies the region 
in which the two drives are comparable or simultaneously weak.

\subsection{Limitations and Future Work}

While our model captures essential kinetic effects and cross-scale dynamics, there are inherent limitations. 
Because the electrons are treated drift-kinetically, parallel electron inertia is retained, but finite-Larmor-radius and cyclotron-frequency electron physics are not resolved. 
Consequently, true electron-scale dissipation channels and high-frequency electromagnetic responses are only partially represented. 
A gyrokinetic electron model would recover finite-Larmor-radius effects but still omit cyclotron dynamics, whereas fully kinetic electrons would be required to capture the complete range of electron-scale processes.

Additionally, we use 2D simulations in slab geometry. While ideal for isolating physics and minimizing computational cost, 3D effects, particularly vortex tilting, reconnection X-line spreading, and turbulence anisotropy, may alter the relative roles of LHDI and KHI. Future 3D hybrid simulations will be required to investigate these effects.

Finally, observational comparisons (e.g., with MMS boundary layer crossings) will be essential to validate the inverse cascade hypothesis and distinguish LHDI-seeded plasmoids from those arising via tearing or vortex-induced reconnection.

\section{Conclusion}
\label{sec:conclusion}

We have investigated the nonlinear formation of plasmoids in 2D, low-$\beta_e$ collisionless plasmas through the competing dynamics of the lower-hybrid drift instability (LHDI) and the Kelvin–Helmholtz instability (KHI), using a hybrid kinetic--gyrokinetic Vlasov simulation framework. Our model combines fully kinetic ions and drift-kinetic electrons, allowing us to capture critical microinstability physics and long-timescale evolution simultaneously.

The key results of this study are as follows:

\begin{itemize}
    \item In thin current sheets with steep density gradients, LHDI rapidly grows and saturates nonlinearly, generating turbulence that undergoes an inverse cascade. This cascade leads to the formation of magnetic islands (plasmoids) without requiring a traditional tearing-mode seed.
    
    \item The inverse cascade from LHDI redistributes energy to meso-scale structures, occupying the scale range where KHI would typically develop. This mechanism can effectively suppress KHI by smoothing out or disrupting the velocity shear layer, especially in low-$\beta_e$ regimes.
    
    \item In cases with broader shear layers or weaker gradients, KHI dominates. The vortices formed by KHI can trap and stretch magnetic field lines, leading to internal reconnection and plasmoid formation inside the rolled-up structures.
    
    \item When LHDI and KHI develop in the same region, their spatial overlap produces additional structures that are not present in isolated cases.
When both instabilities develop, LHDI typically appears along the outer regions of KH vortices. This produces a secondary band of fluctuations, and an intermediate layer often forms where these fluctuations interact with the evolving vortical motion.
The combined activity influences the pattern of reconnection and the associated plasma mixing.
Further examination of these conditions can clarify how the relative strengths and growth rates of the two instabilities determine the resulting transport pathways.
    
    \item Parameter studies reveal that LHDI is favored by higher ion-to-electron mass ratio, lower plasma beta, colder electrons (higher $T_i/T_e$), and sharper density gradients~\citep{Thatikonda2025}. These conditions are relevant to magnetospheric boundary layers, the solar wind, and planetary magnetospheres such as Mercury's~\citep{Dargent2019,Hwang2011}.
\end{itemize}

These findings reveal a novel pathway to plasmoid formation in collisionless reconnection environments, one that is seeded not by large-scale tearing or vortex motion, but by kinetic microinstabilities and their nonlinear evolution. The traditional picture of reconnection driven by large-scale fluid instabilities may need to be revised to account for LHDI-induced turbulence and its cross-scale effects.

From a methodological perspective, the hybrid kinetic--gyrokinetic approach employed here offers a powerful compromise between kinetic fidelity and computational scalability. It enables the study of multi-scale phenomena such as LHDI-driven inverse cascades, turbulence-modified reconnection layers, and dynamic mode competition, all with extended spatial domains and long-time evolution.

Further analysis will benefit from examining how the present two-dimensional behavior connects with the broader range of structures that appear in more realistic configurations.
Future work should extend this analysis to three dimensions, where vortex stretching, oblique LHDI modes, and turbulence anisotropy introduce additional spatial variation in the evolution of the instabilities.
A direct comparison between two- and three-dimensional results will also clarify which features of the system are sensitive to dimensionality and which remain unchanged.
Fully kinetic comparisons, especially in realistic solar wind and magnetospheric contexts, will help validate the suppression and seeding mechanisms discussed here.
Such studies can also determine how these mechanisms operate under conditions that vary in mass ratio, flow geometry, and background plasma parameters.

Overall, this study contributes to a growing body of evidence that microinstabilities like LHDI can significantly alter macroscopic reconnection behavior, not merely as passive turbulence, but as active agents that govern the pathways to plasmoid formation and flux rope evolution in space plasmas.

\section*{Acknowledgments}
The authors gratefully acknowledge support from the Helmholtz Young Investigator
Group grant VH-NG-1239. Computational resources were provided by the
MPCDF Center, Garching. We also extend our gratitude to the Theoretical Physics
Department at Ruhr University Bochum for their collaboration and for providing the
base version of the MUPHY I code, which served as the foundation for ssV development.

\bibliographystyle{jpp}

\bibliography{jpp-instructions}

\end{document}